\documentclass[useAMS,usecolumn,usenatbib,usegraphicx]{mn2e}

\title[H$\alpha$ proto-clusters at $z=2.23$]{An H$\alpha$ search for over-dense regions at $z=2.23$\thanks{Based on observations obtained with the Wide Field CAMera (WFCAM) on the United Kingdom Infrared Telescope (UKIRT), and in part on data collected at Subaru Telescope, which is operated by the National Astronomical Observatory of Japan, and collected at the W.M.\ Keck Observatory, which is operated as a scientific partnership among the California Institute of Technology, the University of California and the National Aeronautics and Space Administration.}}
\author[Y. Matsuda et al.]{
\parbox[t]{\textwidth}{\vspace{-1cm}
Y.\ Matsuda,$^{\! 1}$\thanks{E-mail: yuichi.matsuda@durham.ac.uk} Ian Smail,$^{\! 2}$  J.\ E.\ Geach,$^{\! 3}$  P.\ N.\ Best,$^{\! 4}$  D.\ Sobral,$^{\! 4}$  I.\ Tanaka,$^{\! 5}$  F.\ Nakata,$^{\! 5}$  K.\ Ohta,$^{\! 6}$  J.\ Kurk,$^{\! 7}$  I.\ Iwata,$^{\! 5}$ Rich Bielby,$^{\! 1}$ J.\ L.\ Wardlow,$^{\! 8}$  R.\ G.\ Bower,$^{\! 2}$  N.\ Fanidakis,$^{\! 2}$  R.\ J.\ Ivison,$^{\! 4, 9}$  T.\ Kodama,$^{\! 5}$  T.\ Yamada,$^{\! 10}$  K.\ Mawatari\,$^{\! 10}$ and M. Casali\,$^{\! 11}$} \\\\
$^{1}$ Department of Physics, Durham University, South Road, Durham, DH1 3LE\\
$^{2}$ Institute for Computational Cosmology, Durham University, South Road, Durham, DH1 3LE\\
$^{3}$ Department of Physics, McGill University, 3600 Rue University, Montreal, QC H3A 2T8, Canada\\
$^{4}$ SUPA, Institute for Astronomy, Royal Observatory of Edinburgh, Blackford Hill, Edinburgh EH9 3HJ\\
$^{5}$ Subaru Telescope, National Astronomical Observatory of Japan, 650 North A'ohoku Place Hilo, HI 96720, USA\\
$^{6}$ Department of Astronomy, Kyoto University, Kyoto 606-8502, Japan\\
$^{7}$ Max-Planck-Institut f{\"u}r Extraterrestrische Physik, Postfach 1312, 85741 Garching, Germany\\
$^{8}$ Department of Physics and Astronomy, University of California, Irvine, CA 92697, USA\\
$^{9}$ UK Astronomy Technology Centre, Royal Observatory, Blackford Hill, Edinburgh EH9 3HJ\\
$^{10}$ Astronomical Institute, Graduate School of Science, Tohoku University, Aramaki, Aoba-ku, Sendai 980-8578, Japan\\
$^{11}$ European Southern Observatory, Karl-Schwarzschild-Strasse 2, D-85738 Garching, Germany
}
\begin{document}

\date{Accepted ... ; Received ... ; in original form ...}

\pagerange{\pageref{firstpage}--\pageref{lastpage}} \pubyear{2011}

\maketitle

\label{firstpage}

\begin{abstract}

We present the results of a narrow-band (H$_2$S1, $\lambda_c = 2.121\mu$m, $\delta \lambda =0.021\mu$m) imaging search with WFCAM/UKIRT for H$\alpha$ emitters around several potential signposts of rare ($\sim $\,10$^{-7}$--10$^{-8}$\,Mpc$^{-3}$) over-dense regions at $z=2.23$: an over-density of QSOs (2QZ cluster), a powerful, high-redshift radio galaxy (HzRG), and a  concentration of submillimetre galaxies (SMGs) and optically faint radio galaxies (OFRGs).  In total, we detect 137 narrow-band emitter candidates down to emission-line fluxes of 0.5--1\,$ \times $\,10$^{-16}$\,erg\,s$^{-1}$ cm$^{-2}$, across a total area of 0.56 sq.\,degrees (2.1\,$\times$\,10$^{5}$\,comoving\,Mpc at $z=2.23$) in these fields.  The $BzK$ colours of the emitters suggest that at least 80\% of our sample are likely to be H$\alpha$ emitters (HAEs) at $z=2.23$.  This is one of the largest HAE samples known at $z\ga 2$.  We find modest ($\sim 3 \sigma$) local over-densities of emitters associated with all the three targets.  In the 2QZ cluster field, the emitters show a striking filamentary structure connecting  four of the $z=2.23$ QSOs extending over 30\,Mpc (comoving).  In the HzRG and SMG/OFRG fields, the structures appear to be smaller and seen only in the vicinities of the targets.  The $K$-band magnitudes and the H$\alpha$ equivalent widths of the emitters are weakly correlated with the over-density of the emitters: emitters in over-dense region are more evolved systems compared to those in under-dense regions at $z=2.23$.  We find several examples of extended HAEs in our target fields, including a striking example with a spatial extent of 7.5\,arcsec (60\,kpc at $z=2.23$) in the 2QZ field, suggesting that these are relatively common in high-density regions.  We conclude that narrow-band H$\alpha$ surveys are efficient routes to map over-dense regions at high-$z$ and thus to understand the relation between the growth of galaxies and their surrounding large-scale structures.

\end{abstract}


\section{Introduction}

Local galaxy clusters are characterised by populations of passive, early-type galaxies, whose properties contrast markedly with the star-forming, late-type galaxies found in the surrounding low-density field \citep[e.g.][]{1980ApJ...236..351D}.  The main formation phase of the stars in elliptical galaxies in clusters appears to have occurred at high redshift \citep[probably $z \ga 2$, e.g.][]{1997ApJ...483..582E, 2003ApJ...596L.143B, 2009ApJ...690...42M}, in contrast to the field where most of the star-formation activity occurs at $z \la 2$ \citep[e.g.][]{1995ApJ...455..108L, 2005ApJ...621..673T}.  Hence the evolution of galaxies in clusters appears to be {\it accelerated} relative to that in low-density regions \citep[e.g.][]{2005ApJ...626...44S, 2010A&A...518A..18T, 2010arXiv1012.4860T, 2011arXiv1103.4364H}.  As a result, while the average star-formation rate (SFR) of a galaxy {\it decreases} with increasing local galaxy density in the low-redshift Universe \citep[e.g.][]{2002MNRAS.334..673L, 2003ApJ...584..210G}, this trend should reverse at earlier times: with the SFR {\it increasing} with increasing galaxy density \citep{2007A&A...468...33E, 2010MNRAS.402.1980H, 2010ApJ...719L.126T, 2011MNRAS.tmp...47G}.  Hence the progenitors of massive clusters at high-redshifts (proto-clusters) should be identifiable as over-densities of star-forming galaxies \citep{1998ApJ...492..428S, 2005ApJ...626...44S, 2007A&A...461..823V, 2009MNRAS.400L..66M, 2010MNRAS.403L..54M}.  If the growth of the galaxies is synchronised with that of their super-massive black holes, then populations of active galactic nuclei should also be located in these proto-clusters \citep{2003ApJ...599...86S, 2009ApJ...691..687L, 2010MNRAS.407..846D}.

To identify proto-clusters, the most representative, but time-consuming, technique is to find significant redshift over-densities in large spectroscopic redshift surveys of star-forming galaxies \citep{1998ApJ...492..428S, 2005ApJ...626...44S, 2009ApJ...691..560C, 2009A&A...504..331K}.  A quicker route is to perform such searches around luminous high-redshift sources, such as quasi-stellar objects (QSOs) or powerful high-redshift radio galaxies (HzRGs), where the expectation is that the massive black holes in these galaxies will be hosted by correspondingly massive galaxies which will signpost over-dense regions at high redshifts.  An even more efficient technique is to forego spectroscopy and instead search for concentrations of emission-line galaxies in narrow-band imaging surveys of these regions \citep[e.g.][]{1996Natur.382..231H, 1996Natur.383...45P, 1999AJ....118.2547K, 2000A&A...358L...1K, 2004A&A...428..793K, 2004A&A...428..817K, 2007A&A...461..823V, 2007ApJ...663..765K, 2010arXiv1012.1869T, 2011arXiv1103.4364H}.

For target QSOs or HzRGs at $z>2$,  Ly$\alpha$ is redshifted into the optical and hence most of the narrow-band imaging surveys have targeted Ly$\alpha$ emission.  However, Ly$\alpha$ is far from ideal as it is a resonance line and even a small amount of dust is enough to destroy the line, thus biasing searches against the dusty and perhaps most active galaxies in any structure.  A better choice is to use H$\alpha$, which is less sensitive to dust and also a more accurate tracer of star formation \citep{1998ARA&A..36..189K, 2010MNRAS.402.2017G}.  However, for galaxies at $z>0.4$, H$\alpha$ is redshifted into the near-infrared, and it is only through the recent development of panoramic, near-infrared imagers that narrow-band searches based on H$\alpha$ have become possible \citep[][hereafter G08]{2009MNRAS.398...75S, 2009MNRAS.398L..68S, 2010MNRAS.404.1551S, 2011MNRAS.411..675S, 2008MNRAS.388.1473G}.

We have exploited the wide-field, near-infrared imaging capabilities of WFCAM on UKIRT \citep{2007A&A...467..777C} to carry out an H$\alpha$ imaging survey around several potential signposts of over-dense regions at $z=$\,2.23.  The paper is structured as follows: \S2 describes our target selection, while \S3 details our observations and data reduction and \S4 describes the results derived from these data. Finally, \S5 discusses our results and summarises our main conclusions. We use Vega magnitudes unless otherwise stated and adopt cosmological parameters, $\Omega_{\rm M} = $\,0.27, $\Omega_{\Lambda} =$\,0.73 and $H_0 = $\,73\,km\,s$^{-1}$\,Mpc$^{-1}$. In this cosmology, the Universe at $z=$\,2.23 is 2.9\,Gyr old and 1.0\,arcsec corresponds to a physical length of 8.1\,kpc.

%
%
\begin{figure}
\centering
  \includegraphics[scale=0.46]{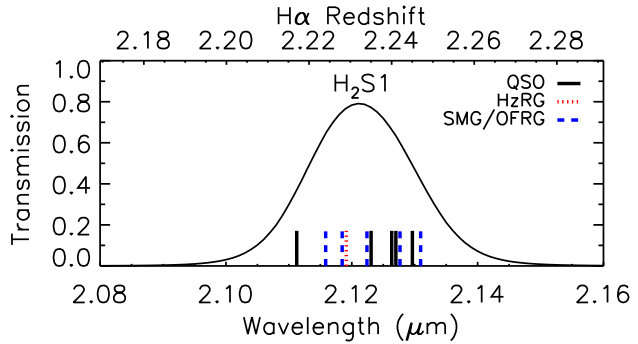}
  \caption{The transmission curves of the H$_2$S1 filter on UKIRT/WFCAM.  We identify the wavelengths for H$\alpha$ emission expected for the redshifts of our various targets: the QSOs in the 2QZ structure, the HzRG MRC\,0200+015 and the SMG/OFRGs in SSA\,13.  The mean redshifts of the three sets of targets are clearly well-matched to the transmission of the filter, allowing us to efficiently survey for H$\alpha$ emitters in any associated structures. }
\end{figure}

%
%
\begin{table*}
\centering
\begin{minipage}{105mm}
  \caption{Summary of targets}
  \begin{tabular}{@{}ccccc@{}}
  \hline
Field & Target & Redshift & Magnitude & Ref.$^a$\\
 & & & \\
\hline
{\bf 2QZ\,cluster} & 2QZ\,J100351.5+001501 & 2.217 & $B=20.43$ & 1 \\
& 2QZ\,J100412.8+001257 & 2.240 & $B=18.57$ & 1, 2 \\
& 2QZ\,J100339.7+002109 & 2.241 & $B=19.58$ & 1, 2 \\
& 2QZ\,J100323.0+000725$^b$ & 2.235 & $B=20.57$  & 1, 2 \\
& 2QZ\,J100204.0+001643$^c$ & 2.245 & $B=20.42$  & 1, 2 \\
\hline
{\bf 0200+015} & NVSS\,J020242+014910 & 2.229  & $H=19.26$ & 3, 4, 5, 6 \\
\hline
{\bf SSA\,13} & SMM\,J131230.92+424051.0 & 2.247  & $K=19.29$  & 7 \\
& SMM\,J131239.14+424155.7 & 2.242 & $K=19.49$  & 7 \\
& RG\,J131207.74+423945.0 & 2.228  & $K=19.36$ & 8 \\
& RG\,J131208.34+424144.4 & 2.234  & $K=19.10$ & 8 \\
& RG\,J131236.05+424044.1 & 2.224  & $K=20.50$ & 8 \\
\hline
\end{tabular}
$^a$ (1) \citet{2004MNRAS.349.1397C}, (2)  \citet{2007AJ....133.2222S} (3) \citet{1981MNRAS.194..693L}, (4) \citet{1997A&A...326..505R}, (5) \citet{1998AJ....115.1693C}, (6) \citet{2003ApJ...598..178I}, (7) \citet{2005ApJ...622..772C}, (8) \citet{2004ApJ...616...71S}.\\
$^b$ This source is just outside of the WFCAM field of view.\\
$^c$ This source is outside of the WFCAM field of view.\\
\end{minipage}
\end{table*}

\section[]{Target field selection}

Our survey uses  the H$_2$S1 narrow-band filter on WFCAM to isolate H$\alpha$ emitters at $z=$\,2.23.  We plot the transmission curve of the narrow-band filter (H$_2$S1, $\lambda_c = 2.121\mu$m, $\delta \lambda =0.021\mu$m) in Figure~1. The redshift range of the H$\alpha$ line corresponding to the 50\% transmission wavelengths of the filter is $z=2.216$--2.248 (equivalent to a comoving length of 43.1\,Mpc along the line of sight).  We therefore searched for targets which could be potential signposts of over-dense regions at $z\sim 2.23$ using the NASA/IPAC Extragalactic Database (NED).\footnote{The NASA/IPAC Extragalactic Database (NED) is operated by the Jet Propulsion Laboratory, California Institute of Technology, under contract with the National Aeronautics and Space Administration.} We selected three targets; an over-density of QSOs, a HzRG, and a concentration of submillimetre galaxies (SMGs) and optically faint radio galaxies (OFRGs). We summarise the targets in Table~1.

\subsection[]{2QZ cluster}

Our first target is a concentration of QSOs selected from the 2dF QSO redshift survey \citep[2QZ,][]{2001MNRAS.322L..29C}.  There now appears to be some consensus that on average, over-densities of galaxies are present around typical QSOs (both radio-loud and quiet) at $z<2$ \citep[e.g.][]{1991ApJ...371...49E, 1998ApJ...507..558H}.  However, using {\it over-densities} of QSOs should be a much clearer marker of structures at high redshift \citep{1991MNRAS.249..218C}.  We searched the whole equatorial region from the 2QZ survey for regions with more than four QSOs at $z=$\,2.216--2.248 in a 1-degree diameter field.  There are 285 QSOs at $z=2.216$--2.248 in the 2QZ survey area of 289.6 deg$^2$ (or $1.1 \times 10^8$ comoving\,Mpc).  We found only one structure satisfying the criteria in this volume indicating a volume density of any associated structure of $\sim $\,10$^{-8}$ comoving\,Mpc$^{-3}$.  We refer to this target as the 2QZ cluster, it contains five QSOs at $z=$\,2.217, 2.235, 2.240, 2.241 and 2.245 (see Figure~1 and Table~1).\footnote{Four of the 5 QSOs are also listed in the SDSS QSO catalog \citep{2007AJ....133.2222S}}  Four out of the five QSOs are even more strongly clustered in a $15 \times 15$ arcmin$^2$ region, with three of these QSOs falling within the field of view of a single WFCAM chip, with the fourth located just outside the field of view.

\subsection[]{0200+015}

Our second target is a HzRG: MRC\,0200+015.  The number density of HzRGs at $z=$\,2--5 is a few times $10^{-8}$\,Mpc$^{-3}$ \citep{2007A&ARv..15...67M}, and thus HzRGs are quite rare.  There is growing evidence that a significant fraction of HzRGs reside in over-dense environments \citep{2000A&A...358L...1K, 2000A&A...361L..25P, 2003Natur.425..264S, 2003ApJ...599...86S, 2004Natur.427...47M, 2006MNRAS.371..577K, 2007MNRAS.377.1717K, 2007A&A...461..823V, 2010arXiv1012.1869T, 2011arXiv1103.4364H}.  However, \citet{2003MNRAS.343....1B} show that HzRG at $z \sim $\,1.5--2 are found in a very wide range of environments, from essentially no over-density, through a small-scale central over-density to larger scale over-densities.  Using NED, we identified a HzRG, MRC\,0200+015 (or NVSS\,J020242+014910) at $z=2.229$ \citep[hereafter 0200+015,][]{1981MNRAS.194..693L, 1997A&A...326..505R, 1998AJ....115.1693C}.  This HzRG field was observed using H$\alpha$ imaging by \citet{2000A&A...362..509V}.  They imaged an area of just 6.37 arcmin$^2$ down to a 3\,$\sigma$ flux limit of 1.0\,$\times$\,10$^{-16}$\,erg\,s$^{-1}$\,cm$^{-2}$ for H$\alpha$ emitters in redshift range of $z=$\,2.19--2.26.  Although they detected the HzRG as an H$\alpha$ emitter, they did not identify any additional H$\alpha$ emitter candidates around the HzRG.  A single WFCAM chip has a field of view $\sim 30$ times larger than the area surveyed by \citet{2000A&A...362..509V}, giving us the first opportunity to conclusively search for a structure around this HzRG.

\subsection[]{SSA\,13}

Our final target is a potential concentration of massive starburst galaxies.  From redshift surveys of SMGs and OFRGs, a significant redshift spike was discovered at $z=$\,2.224--2.247 ($\Delta z=$\,0.023) in the Small Selected Area 13 (SSA\,13) field \citep{2004ApJ...616...71S, 2005ApJ...622..772C}.  This spike contains two SMGs at $z=$\,2.242, 2.247 and three OFRGs at $z=$\,2.224, 2.228, and 2.234 (see Figure~1 and Table~1).  This spike is one of the most prominent structures in their survey volume of a few times 10$^6$ comoving\,Mpc, suggesting a structure with a number density of a few times $10^{-7}$\,Mpc$^{-3}$ \citep[c.f.][]{2009ApJ...691..560C}.

%
%
\begin{table*}
\centering
\begin{minipage}{180mm}
  \caption{Summary of observations and data}
  \begin{tabular}{@{}cccccccccc@{}}
  \hline
Field &  Coordinate (J2000) & Date & Filter & Chip & Exp time & Depth$^a$ & FWHM & Area & Number density$^b$ \\
 &  (h:m:s) (d:m:s)  & (mm/yyyy) & & & (ks) & (mag) & (arcsec) & (arcmin$^2$) & (arcmin$^{-2}$)\\
\hline
{\bf 2QZ\,cluster} & 10:03:51.0 +00:15:09 & 02/2010 & H$_2$S1 & 1 & 34.44 & 19.9 & 0.9 & 169 & 8.1 (H$_2$S1$\le$19.9)\\
 & 10:05:37.2 +00:15:03 & & & 2 & 34.44 & 20.0 & 0.9 & 171 & 9.6 (H$_2$S1$\le$19.9)\\
 & 10:05:37.8 +00:41:22 & & & 3 & 34.44 & 19.9 & 0.9 & 172 & 10.1 (H$_2$S1$\le$19.9)\\
 & 10:03:51.5 +00:41:33 & & & 4 & 34.44 & 20.1 & 0.9 & 171 & 8.6 (H$_2$S1$\le$19.9)\\
 & 10:03:51.0 +00:15:09 & 02-03/2010 &  $K$  & 1 & 4.48 & 20.4 & 0.7 & 169 & 8.2 ($K$$\le$19.9)\\
 & 10:05:37.2 +00:15:03 & & & 2 & 4.48 & 20.3 & 0.8 & 171 & 10.0 ($K$$\le$19.9)\\
 & 10:05:37.8 +00:41:22 & & & 3 & 4.48 & 20.4 & 0.8 & 172 & 10.3 ($K$$\le$19.9)\\
 & 10:03:51.5 +00:41:33 & & & 4 & 4.48 & 20.3 & 0.8 & 171 & 8.8 ($K$$\le$19.9)\\
\hline
{\bf 0200+015} & 02:02:42.6 +01:50:54 & 10/2007 & H$_2$S1  & 3 & 13.44 & 19.6 & 0.9 & 148 & 8.3 (H$_2$S1$\le$19.5)\\
 & 02:00:56.0 +01:24:38 & & & 1 & 13.44 & 19.6 & 1.0 & 172  & 8.1 (H$_2$S1$\le$19.5)\\
 & 02:02:42.2 +01:24:35 & & & 2 & 13.44 & 19.7 & 1.1 & 169 & 7.2 (H$_2$S1$\le$19.5)\\
 & 02:00:56.1 +01:51:02 & & & 4 & 13.44 & 19.5 & 1.0 & 174 & 6.5 (H$_2$S1$\le$19.5)\\
 & 02:02:42.6 +01:50:54 & 10/2007 & $K$ & 3 & 1.335 & 20.1 & 1.0  & 148 & 8.0  ($K$$\le$19.5)\\
 & 02:00:56.0 +01:24:38 & & & 1 & 1.23 & 20.1 & 1.1  & 172 & 8.0  ($K$$\le$19.5)\\
 & 02:02:42.2 +01:24:35 & & & 2 & 1.25 & 20.1 & 1.0  & 169 & 7.4  ($K$$\le$19.5)\\
 & 02:00:56.1 +01:51:02 & & & 4 & 1.365 & 20.0 & 1.2  & 174 & 6.5  ($K$$\le$19.5)\\
\hline
{\bf SSA\,13} & 13:12:34.1 +42:40:43 & 05/2006 & H$_2$S1 & 1 & 14.76 & 19.0 & 0.9 & 146 & 4.2  (H$_2$S1$\le$19.0)\\
 & 13:14:58.5 +42:40:49 & & & 2 & 14.32 & 19.1 & 0.9 & 174 & 5.1  (H$_2$S1$\le$19.0)\\
 & 13:14:58.2 +43:07:03 & & & 3 & 14.56 & 19.1 & 0.9 & 174 & 4.9  (H$_2$S1$\le$19.0)\\
 & 13:12:33.2 +43:07:08 & & & 4 & 14.96 & 19.1 & 0.9 & 171 & 4.6  (H$_2$S1$\le$19.0)\\
 & 13:12:34.1 +42:40:43 & 05/2006 & $K$ & 1 & 0.625 & 19.7 & 1.0 & 147 & 4.3 ($K$$\le$19.0)\\
 & 13:14:58.5 +42:40:49 & & & 2 & 0.615 & 19.7 & 0.9  & 174 & 5.0  ($K$$\le$19.0)\\
 & 13:14:58.2 +43:07:03 & & & 3 & 0.63 & 19.8 & 0.9  & 174 & 4.8 ($K$$\le$19.0)\\
 & 13:12:33.2 +43:07:08 & & & 4 & 0.62 & 19.5 & 0.9  & 171 & 4.7  ($K$$\le$19.0)\\
\hline
\end{tabular}
$^a$The 5\,$\sigma$ limiting magnitude based on photometry with a 3-arcsec diameter aperture.\\
$^b$The number density of H$_2$S1 detected sources.\\
\end{minipage}
\end{table*}

\section[]{Observations and Data Reduction}

\subsection{WFCAM Observations}

The three target fields were observed between 2006 May and 2010 April with WFCAM on UKIRT, using the $K$-band and H$_2$S1 filters. We summarise the observations in Table~2.  WFCAM has four $13.7 \times 13.7$\,arcmin$^2$ chips offset by 20\,arcmin (a comoving separation of 32\,Mpc at $z=2.23$). For our observations we place the target on one of the chips (Chip 1 for the 2QZ cluster and SSA\,13, Chip 3 for 0200+015), with the other three chips providing control fields to derive the mean density of emitters $\sim 40$--60\,Mpc (comoving) away from the target.  To help with cosmic ray rejection over the relatively long narrow-band exposures (40--60\,s), we used the NDR (Non Destructive Read) mode, whereas CDS (Correlated Double Sampling) mode was used for the shorter broad-band exposures. To improve sampling of the PSF with the 0.4\,arcsec pixels of WFCAM, the narrow-band frames were microstepped in a $2 \times 2$ grid with 1.2\,arcsec offsets at each position, following a 14-point jitter sequence.  The seeing in our observations varied between 0.7--1.2\,arcsec FWHM.

The data reduction was carried out in the same manner as for the HiZELS survey \citep[G08;][]{2009MNRAS.398...75S, 2009MNRAS.398L..68S, 2010MNRAS.404.1551S, 2011MNRAS.411..675S}.  We flatfield a given image using a normalised median combination of the 13 remaining frames from the same sequence, taking care to mask-out bright sources in each frame.  A world coordinate system is then automatically fit to each frame by querying the USNO A2.0 catalogue, on average returning $\sim 100$ sources to derive the astrometric fit.  Frames are aligned and co-added with {\sc swarp} \citep{2002ASPC..281..228B}.  Both $K$-band and H$_2$S1 magnitudes were calibrated by matching $K=$\,10--15 stars from the 2MASS All-Sky Catalogue of Point Sources \citep{2003tmc..book.....C} which are unsaturated in our frames.  The magnitudes were not corrected for Galactic extinction, because the extinction is negligible in these bands \citep[$\la 0.01$ mag,][]{1998ApJ...500..525S}.

The combined images were aligned and smoothed with Gaussian kernels to ensure that the final images in each field have the same seeing  (FWHM\,=\,0.9--1.2\,arcsec). The size of each chip analyzed here is 13.2\,$ \times$\,13.2 arcmin$^2$ after removal of low S/N regions near the edge.  We also masked out halos and cross-talk residuals of the bright stars ($K<$\,15).  The resultant total effective area of each chip is $\sim $\,150--170 arcmin$^2$ (corresponding to a comoving volume of $\sim $\,1.6--1.8\,$\times $\,10$^4$\,Mpc$^3$ for H$\alpha$ emitters at $z=2.23$ in the H$_2$S1 filter). 

Source detection and photometry were performed using {\sc SExtractor} version 2.5.0 \citep{1996A&AS..117..393B}. The source detections were made on the H$_2$S1 image.  We detected sources with five connected pixels above 1.0--1.5\,$\sigma$ of the sky noise.  Each WFCAM chip has four amplifiers, we added small ($< \pm $\,0.1 mag) shifts to the narrow-band zero point magnitudes to make the median $K-NB$ colour zero for each region with the same amplifier in each chip.  The magnitudes and colours are measured for each source in a 3\,arcsec diameter aperture.  In Table~2, we give the number densities of H$_2$S1 detected sources down to the 5\,$\sigma$ limit in each field.  We confirmed that these are consistent with $K$-band number counts of galaxies in UKIDSS/DXS by \citet{2011MNRAS.410..241K}.  The number densities in each chip in each field agree within $\pm$\,12\%, and the variations are also consistent with those seen in the UKIDSS/DXS data (Jae-Woo Kim, private communication).  The agreement suggest that our photometric calibration is reliable for each field.

%
%
\begin{figure*}
\centering
  \includegraphics[scale=1.0]{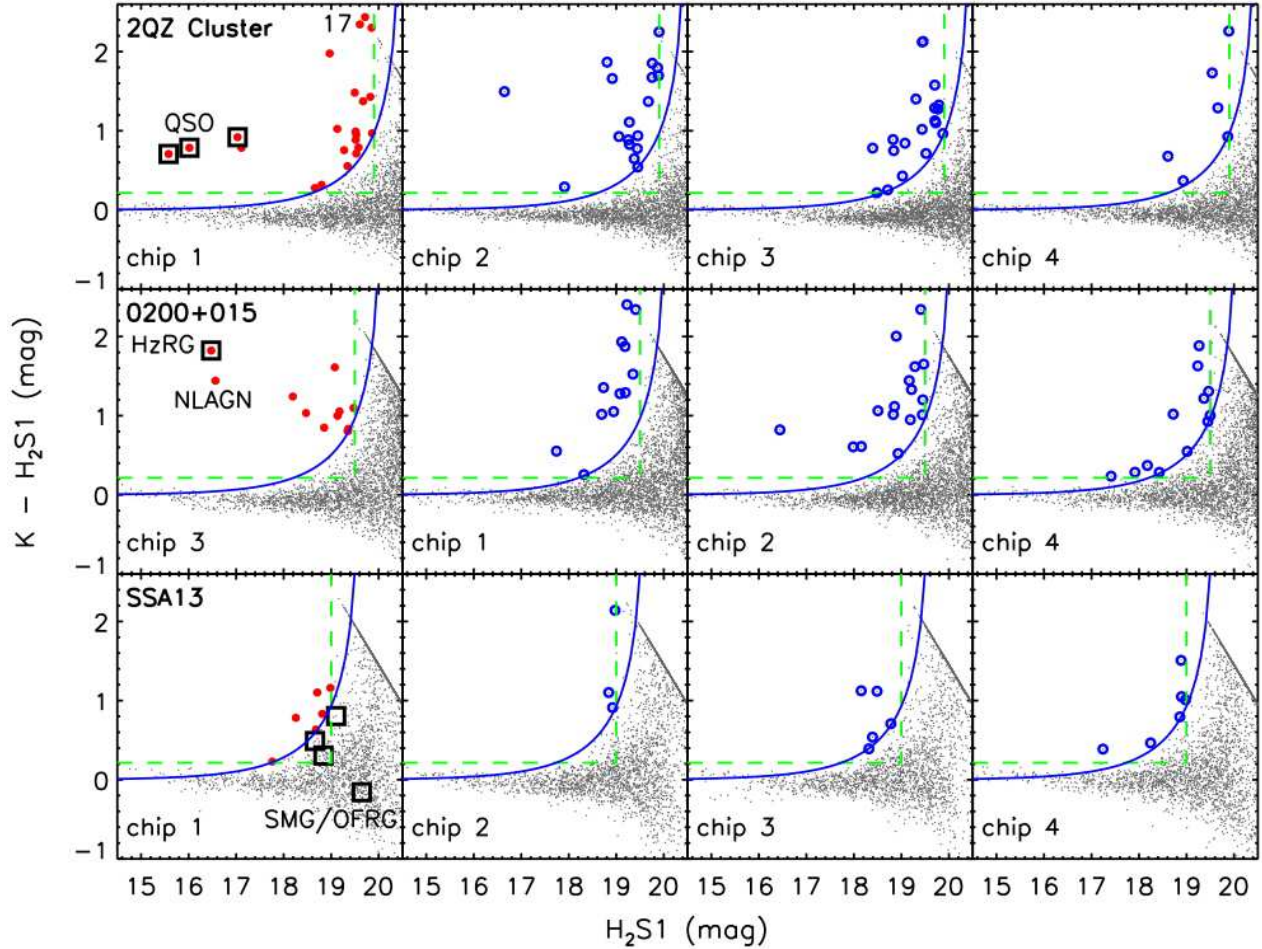}
  \caption{Colour-magnitude plots showing ${\rm H}_2{\rm S1}$ versus $K-{\rm H}_2{\rm S1}$ for the H$_2$S1-detected sources in our three survey fields.  The  filled and  open circles identify the emitter candidates in the target and control fields, respectively (see Appendix A).  The squares indicate the target QSOs, HzRG, and SMG/OFRGs.  The dashed lines are the colour cuts corresponding to $K-{\rm H}_2{\rm S1}=$\,0.215  ($EW_{\rm obs}\ge$\,50\AA) and ${\rm H}_2{\rm S1}>$\,5\,$\sigma$, used to select line emitters in our analysis. The solid curves show the expected 2.5\,$\sigma$ error limit in the colour of a $f_{\nu}$ constant source. We also identify HAE17  in the 2QZ cluster field, which is surrounded by extended H$\alpha$ nebula, and the spectroscopically confirmed narrow-line AGN (NLAGN) at $z=$\,2.235 in the 0200+015 field.  All magnitudes and colours are measured with 3\,arcsec diameter apertures.  Comparison of the numbers of sources in the various fields suggests that no field has an over-density of emitters on scales of a WFCAM chip.}
\end{figure*}

\subsection{Supporting Observations}

\subsubsection{SCAM Observations} To check contamination in our emitter sample from foreground and background line emitters, we obtained $B$ and $z'$-band images of the target fields with Subaru/Suprime-Cam \citep{2002PASJ...54..833M}.  We observed the 2QZ cluster field in 2009 November and the 0200+015 field in 2010 November.  For the SSA\,13 field, we used archival data.  All the images were a single pointing of Suprime-Cam, covering only one chip of the WFCAM observations.  The exposure times were 0.9--3.0\,ks for the $B$-band and 0.9--1.8\,ks for $z'$-band, respectively.  The data were reduced using {\sc sdfred} \citep{2002AJ....123...66Y, 2004ApJ...611..660O}.  For photometric calibration, we used the photometric standard stars in SA\,101 field \citep{1992AJ....104..340L, 2002AJ....123.2121S} and SDSS $z'$-band images for the 2QZ cluster and SSA\,13 fields.  We corrected the magnitudes using the Galactic extinction map of \citet{1998ApJ...500..525S}.  The seeing of the stacked images are 0.9--1.2 arcsec for $B$-band and 0.6--0.8 arcsec for $z'$-band, respectively.  The 1\,$\sigma$ limiting AB magnitudes derived with 3\,arcsec diameter aperture photometry are 27.0--27.6 ABmag for $B$-band and 25.6--26.0 ABmag for $z'$-band, respectively.

\subsubsection{MOIRCS Observations} The 0200+015 field was also observed with the $K_s$-band and H$_2$S1 filters using Subaru/MOIRCS \citep{2008PASJ...60.1347S} in 2007 August as part of engineering tests (performed by IT).  The H$_2$S1 filter on MOIRCS has $\lambda_c = $\,2.116$\mu$m and $\delta \lambda =$\,0.021$\mu$m. The H$\alpha$ redshift range covered with the H$_2$S1 filter is $z=$\,2.208--2.240, which is only $\Delta z=$\,0.008 smaller than that covered with the H$_2$S1 filter on WFCAM.  As one of the two chips had problems, we used only one chip. The exposure times were 1.17\,ks for $K_s$-band and 1.44\,ks for ${\rm H}_2{\rm S1}$, respectively.  We reduced the data using the MOIRCS imaging data reduction pipeline \citep[{\sc mcsred},][]{2010arXiv1012.1869T}.  For photometric calibration, we used the 2MASS All-Sky Catalogue of Point Sources \citep{2003tmc..book.....C}. The seeing of the stacked images are 0.48\,arcsec for $K_s$-band and 0.42\,arcsec for ${\rm H}_2{\rm S1}$. The survey area is 12.4\,arcmin$^2$ after masking low S/N regions. The 1\,$\sigma$ limiting magnitudes calculated with a 0.84\,arcsec diameter aperture are 22.3\,mag for $K_s$-band and 21.0\,mag for H$_2$S1.  The deeper, higher resolution images can be used as a test of the completeness of our WFCAM observations and to examine the H$\alpha$ morphology in the central 4\,$ \times$\,3.5\,arcmin$^2$ part of the 0200+015 field.  These data will be discussed these further in I.\ Tanaka et al.\ (in preparation).

\subsubsection{LRIS Observations} In addition, we carried out a long-slit spectroscopic observation of a bright emitter candidate in the 0200+015 field with Keck/LRIS \citep{1995PASP..107..375O} in 2010 September.  As the red-side CCD had problems, we used only the blue arm.  We used the 400/3400 grism and a 1.0\,arcsec slit, yielding spectral coverage across $\sim $\,3500--5700\,\AA, at a spectral resolution of FWHM\,$\sim$\,7\,\AA\ or $\sim $\,600\,km\,s$^{-1}$.  The exposure time was 0.9\,ks and we reduced the data with standard {\sc iraf} tasks. For wavelength calibration, we used arc lamp spectra with Hg, Cd, and Zn lines, giving a wavelength calibration with an rms $\sim $\,0.6\,\AA.  We also confirmed that there is no overall shift ($\la $\,0.2\,\AA) for the wavelength calibration using the strong [O{\sc i}] sky emission at 5577.3\,\AA.

%
%
\begin{figure*}
\centering
  \includegraphics[scale=1.0]{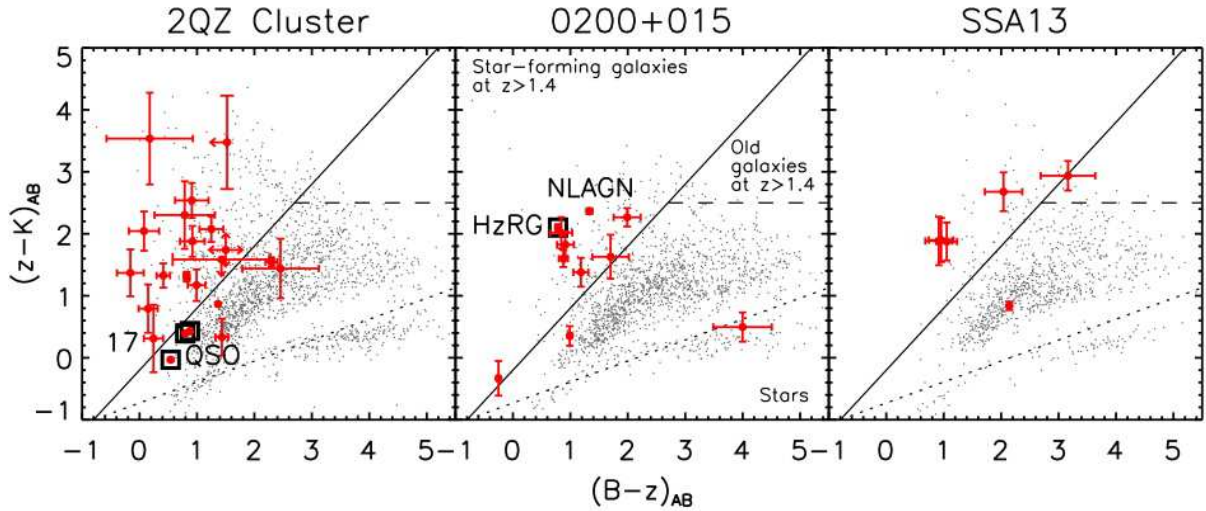}
  \caption{BzK diagram for emitter candidates in the target fields.  At least 80\% of the emitter candidates satisfy the $BzK$ criteria, $BzK\equiv (z-K)_{\rm AB} - (B-z)_{\rm AB} \ge -$0.2 or $(z-K)_{\rm AB} >$\,2.5 \citep{2004ApJ...617..746D}, indicating that these are likely to be H$\alpha$ emitters at $z=$\,2.23.  This plot suggests that the contamination rate from potential foreground and background line emitters in our emitter sample is $\la $\,20\%.  The squares indicate the target QSOs, HzRG.  The HAE17 with an extended emission-line nebula in the 2QZ cluster field and spectroscopically confirmed narrow-line AGN (NLAGN) at $z=$\,2.235 in the 0200+015 field are also marked.}
\end{figure*}

%
%
\begin{table}
\centering
\begin{minipage}{75mm}
  \caption{Numbers and over-density of emitter candidates on WFCAM chip scales.}
  \begin{tabular}{@{}cccccc@{}}
  \hline
Field & Chip & N & n & $\delta^a$ \\
 & & & (arcmin$^{-2}$) &  \\
\hline
{\bf 2QZ cluster} &  1  & 22 & $0.13\pm0.03$ & $0.5\pm0.4$\\
Control & 2,3,4  & 44 & $0.09\pm0.01$ & $0.0\pm0.2$\\
 & 2  & 18 & $0.11\pm0.02$ & $0.2\pm0.3$\\ 
 & 3  & 20 & $0.12\pm0.03$ & $0.4\pm0.4$\\
 & 4  & 6 & $0.04\pm0.01$ & $-0.6\pm0.2$\\
\hline
{\bf 0200+015}  & 3 & 11 & $0.07\pm0.02$ & $0.0\pm0.3$\\
Control  & 1,2,4  & 40 & $0.08\pm0.01$ & $0.0\pm0.2$\\
 & 1 & 12 & $0.07\pm0.02$ & $-0.1\pm0.3$\\
 & 2 & 16 & $0.09\pm0.02$ & $0.2\pm0.4$\\
 & 4 & 12 & $0.07\pm0.02$ & $-0.1\pm0.3$\\
\hline
{\bf SSA\,13} &  1 & 6 & $0.04\pm0.02$ & $0.5\pm0.7$\\
Control & 2,3,4  & 14 & $0.03\pm0.01$ & $0.0\pm0.3$\\
 &  2 & 3 & $0.02\pm0.01$ & $-0.4\pm0.4$\\
 &  3 & 5 & $0.03\pm0.01$ & $0.1\pm0.6$\\
 &  4 & 6 & $0.04\pm0.01$ & $0.3\pm0.6$\\
\hline
\end{tabular}
$^a$ $\delta\equiv(n-\bar{n})/\bar{n}$, where $\bar{n}$ is the mean density of emitter candidates derived from the three chips surrounding each target field.\\
\end{minipage}
\end{table}

%
%
\begin{figure*}
\centering
  \includegraphics[scale=0.53]{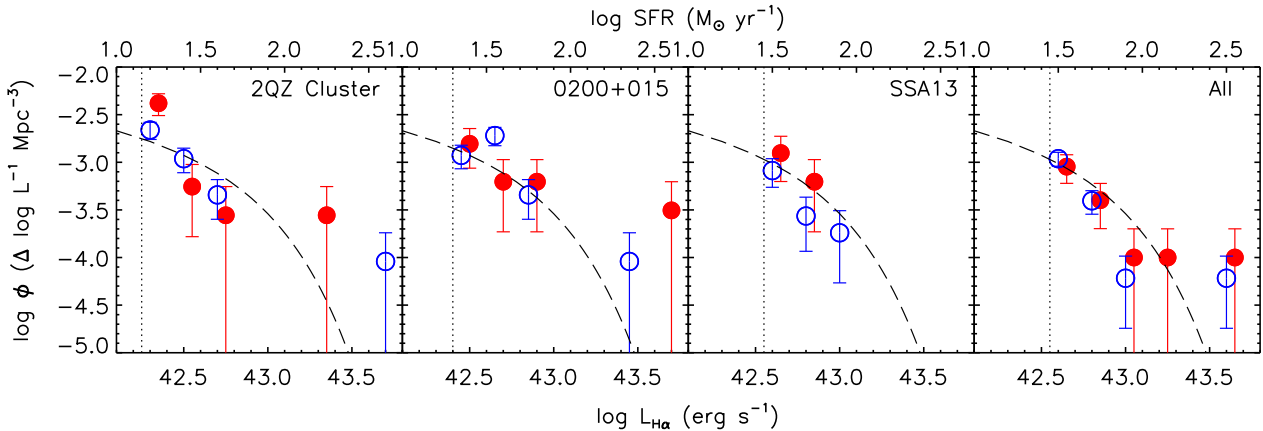}
  \caption{The H$\alpha$ luminosity function (LF) in the target and control fields.  The LFs in the target (filled symbol) and control fields (open symbol) appear to be consistent with that of the blank field.  We exclude the QSOs and HzRG from the emitter sample used to derive the LF.  As the contamination rate is $\la $\,20\% in our emitter sample in the target fields, we assume all the emitter candidates are H$\alpha$ emitters in this plot.  The dashed line shows the blank field H$\alpha$ luminosity function at $z=$\,2.23 from \citet{2008MNRAS.388.1473G}.  The dotted lines represent the detection limits of our emitter sample.  The data points in the control fields are slightly shifted ($-0.1$ dex) for display purposes.}
\end{figure*}

%
%
\begin{figure*}
\centering
  \includegraphics[scale=.6]{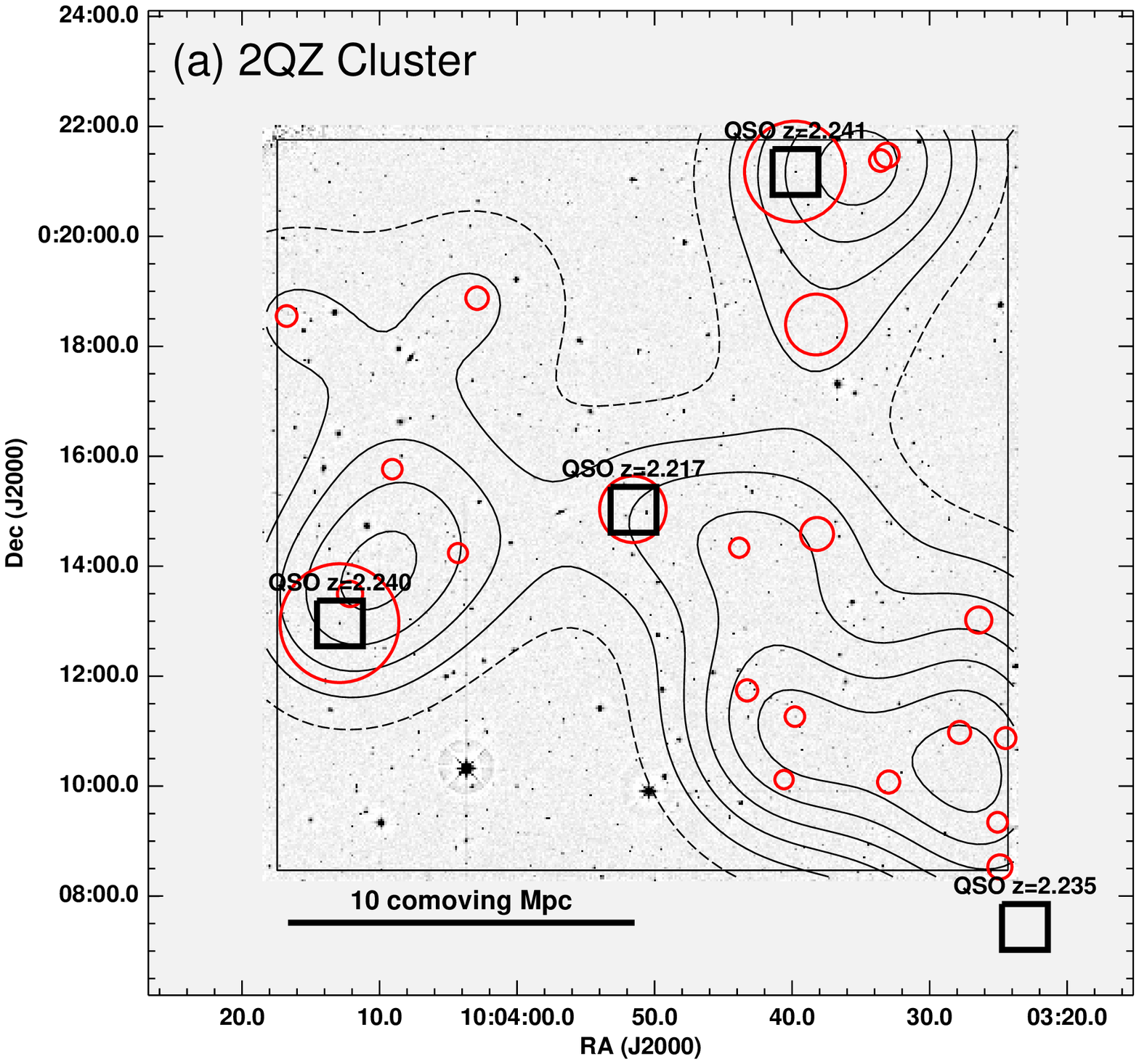}
  \includegraphics[scale=0.9]{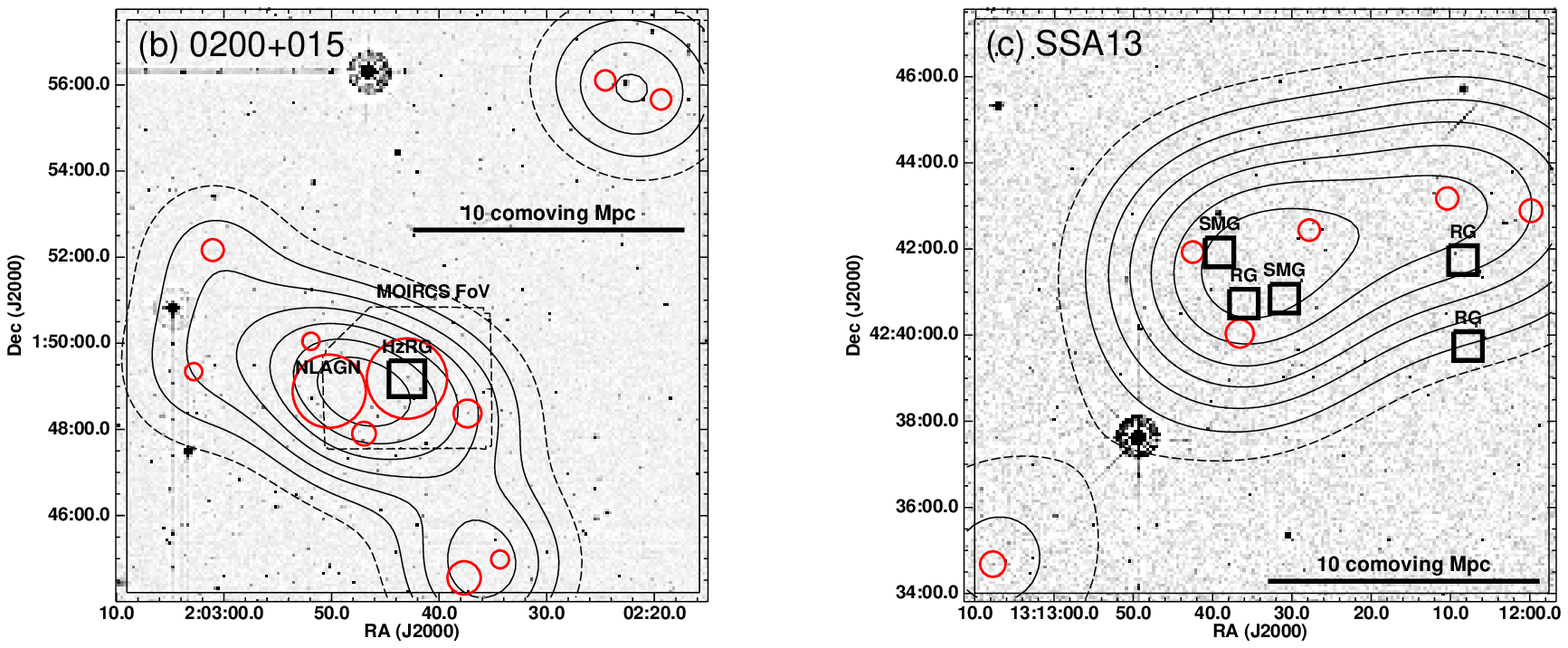}
  \caption{The sky distribution and smoothed density map of emitter candidates at $z\sim$\,2.23 in the (a) 2QZ cluster field, (b) 0200+015 field, and (c) SSA\,13 field.  In the 2QZ cluster field, we see a filamentary structure which appears to connect up the four QSOs at $z=$\,2.23 in this region.  The contours show deviations of local emitter densities from the average densities, from -0.5 to 3\,$\sigma$ with 0.5\,$\sigma$ intervals.  In the 0200+015 field, we see an over-density centred on the HzRG.  The region surrounded by the dashed line is the area surveyed with MOIRCS.  Our MOIRCS observations confirm the reality of the four narrow-band excesses sources in this region.  In the SSA\,13 field, we see a modest over-density near to the SMG/OFRG structure.  The large solid boxes show the survey areas of WFCAM overlaid on the WFCAM H$_2$S1 images.  The squares indicate the positions of the QSOs, HzRG, SMGs and OFRGs.  The circles indicate the emitter candidates and the size is proportional to the emission-line flux.  The scale bars show the angular size corresponding to 10 comoving\,Mpc (3.1\,Mpc in physical scale) at $z=$\,2.23.}
\end{figure*}

\section[]{Analysis and Results}

We show in  Figure~2 the colour--magnitude plots for the ${\rm H}_2{\rm S1}$-detected sample.  From the ${\rm H}_2{\rm S1}$-detected sources with ${\rm H}_2{\rm S1} \ge $\,5\,$\sigma$ in each field, we select emitter candidates with the following criteria: 
\begin{enumerate}
\renewcommand{\theenumi}{(\arabic{enumi})}
\item $K-{\rm H}_2{\rm S1} \ge$\, 0.215 (EW$_{\rm obs}\ge $\,50 \AA ),
\item $\Sigma \ge $\,2.5, 
\end{enumerate}
where $\Sigma$ is the ratio between the H$_2$S1 excess and the uncertainty in the $K-{\rm H}_2{\rm S1}$ colour based on photometric errors of both $K$ and H$_2$S1 for sources with a constant $f_{\nu}$ spectra.  These criteria are the same as used in G08.  One slight difference from G08 is that we don't correct the $K$-band magnitudes using the $z'-K$ colour for the emitter selection, because we don't have $z'$-band images in our control fields.  As we show below, this does not appear to adversely effect the purity of our narrow-band excess sample.  We note that in each field, the limiting magnitudes between the chips are slightly different and so we use the shallowest 5\,$\sigma$ limits and significance curves for the colour  cut for both the targets and the surrounding control fields to ensure a fair comparison of the number density  in each field.  The colour  cuts correspond to flux limits of $\sim $\,0.5, $\sim$\,0.7 and $\sim $\,1.0\,$\times $\,10$^{-16}$\,erg\,s$^{-1}$ cm$^{-2}$ for the 2QZ cluster,  0200+015 and SSA\,13 fields respectively.

In total we detect 137 emitter candidates over a combined area of 0.56\,sq.\ degrees (2.1\,$ \times$\,10$^{5}$\,comoving\,Mpc$^3$ at $z=$\,2.23) in the three fields.  A flux limit of 0.5\,$ \times$\,10$^{-16}$\,erg\,s$^{-1}$\,cm$^{-2}$ corresponds to a star-formation rate (attenuation-uncorrected) of SFR $\sim $\,14\,M$_\odot$\,yr$^{-1}$ using the calibration of \citet{1998ARA&A..36..189K} or SFR $\sim $\,70\,M$_\odot$\,yr$^{-1}$ (attenuation-corrected) using the reddening estimates from \citet{2010MNRAS.402.2017G}.  We listed the resulting catalogue of emitter candidates in the tables in Appendix A.

We have used the Suprime-Cam optical imaging to investigate the contamination  from potential foreground and background line emitters (e.g., Pa$\alpha$ emitters at $z=$\,0.13, Pa$\beta$ emitters at $z=$\,0.65 and [O{\sc iii}]5007 emitters at $z=$\,3.24).  We show in Figure~3 the $B-z'$ and $z'-K$ colour-colour plot for emitter candidates in the target fields.  This $BzK$ colour-colour plane allows us to isolate $z\sim $\,1.4--2.5 galaxies \citep{2004ApJ...617..746D}, which are likely to be H$\alpha$ emitters (HAEs), from any foreground or background contamination (see G08).  Although the $BzK$ criteria, $BzK\equiv (z-K)_{\rm AB} - (B-z)_{\rm AB} \ge -$0.2 or $(z-K)_{\rm AB} >$\,2.5, were determined using a $BzK$ filter set from different instruments, it has been confirmed that the same criteria can also be used for Suprime-Cam $B$ and $z'$, and WFCAM $K$-bands to select $z\sim $\,1.4--2.5 galaxies \citep[e.g.][]{2009ApJ...691..140H}.  We find that $\ga $\,80\% of the emitter candidates satisfy the $BzK$ criteria, indicating that these are likely to be $z\sim $\,2.23 HAEs.  Note that the three QSOs in the 2QZ cluster field are slightly outside the $BzK$ colour-selection region ($BzK \sim $\,0.0--0.2 mag), even though they are spectroscopically confirmed at $z \sim $\,2.23.  As the $BzK$ analysis was designed for galaxies, and not for AGNs, this it is perhaps not that surprising.   

The apparent contamination rate in our candidate emitter sample, $\sim 20$\,\% is similar to the $\sim 10-20$\,\% rates in other H$\alpha$ emitter surveys for protoclusters \citep{2010arXiv1012.1869T,  2011arXiv1103.4364H}.  Although our contamination rate seems to be lower than the $\sim 50-70$\,\% rates in other H$\alpha$ emitter surveys in blank fields \citep[G08;][]{2010A&A...509L...5H, 2011arXiv1103.4364H}, it is similar to the $\sim 30$\,\% rate for the sub-sample of \citet{2010A&A...509L...5H}'s emitters with the similar flux range to our sample.  As we show below, the resulting H$\alpha$ luminosity functions of our emitters (in both target and control fields) and G08's HAEs are consistent,  supporting the low contamination in our HAE sample.  We note that for the following comparison of the number density between the target  and  control fields, we have to use the full sample of emitter candidates (i.e., before applying the $BzK$ criteria) because we lack optical images in our control fields.  However, due to our low contamination rate this only slightly affect the significance of the over-densities we find in these fields.

In Table~3, we summarise the number, number density, and over-density of the emitter candidates in each field.  We derive the mean emitter densities using the surrounding control fields for each target field.  Our analysis of the number densities suggests that there is no significant excess of emitters in any of the three target fields on the scale of the WFCAM chips (22\,comoving\,Mpc at $z=$\,2.23).  Note that in the 2QZ cluster field, the number of emitter candidates on chip 4 appears to be lower than on the other chips.  However, we have confirmed that across all fields the total number of all H$_2$S1-detected sources (not just those showing excess emission in the H$_2$S1 filter) on chip 4 is not different with those in the other chips, as shown in Table~2.  Thus, the lower numbers of the emitter candidates on chip 4 in the 2QZ cluster field are likely to be real (i.e., a void of emitters). 

We compare the H$\alpha$ luminosity functions (LF) of the emitter candidates in the target and control fields in Figure~4.  All the LFs appear to be consistent with the blank field LF of $z=$\,2.23 HAEs from G08.  We could not find any clear difference between the shapes of LFs in the target and control fields.  In this comparison, we have excluded the three QSOs and HzRG from our emitter sample.

To search for over-densities on scales smaller than the chip field-of-view, we show in Figure~5 the sky distribution and smoothed density map of emitter candidates in the target fields.  The surface density maps are generated with a gaussian smoothing kernel with a  size chosen to match the median distance between the nearest-neighbour emitters in the control fields, $\sigma=$\,1.4\,arcmin for the 2QZ cluster and 0200+015, and $\sigma=$\,2.3\,arcmin for SSA\,13.  In these maps, we can see modest over-densities ($\sim $\,3\,$\sigma$ deviations from the average densities) of emitters within all  three target fields. In the 2QZ cluster field, the emitters appear to have a filamentary structure connecting the four QSOs, while  in the HzRG and SMG/OFRG fields, the structures appear to be smaller and only seen in the vicinities of the targets. We discuss these structures below, but we first note that the estimated significance of these over-densities may be conservative, because the potential contamination will be unclustered and so should slightly decrease the significance of any real over-densities, although again we stress that the contamination rate in our emitter sample is expected to be quite low ($\sim $\,20\%) so this should not be a large effect.\\

\noindent{\bf 2QZ cluster:}  As can be seen in Figure\,5, the over-density in the 2QZ cluster field comprises an apparently filamentary structure connecting at least  three of the QSOs at $z=$\,2.23, as well as an extension of the structure towards the QSO just outside of the field of view to the south west.  The three target QSOs lie in weak over-dense regions ($\delta \sim $\,0--1) rather than the local density peaks. The structure around the QSOs contains a 2.9\,$\sigma$ density peak from the average density derived in the control field, and it is the second highest peak in the full density map of the 2QZ cluster and the surrounding three control fields (see Figure~B1 in Appendix).  The local emitter density of this peak is 3.7 times higher than the average density.    The highest density peak is located in the west edge of the chip 2.  There is a very bright, point-source emitter near to this density peak.  This bright emitter may be another luminous AGN at $z=$\,2.23 as the emission-line luminosity is similar to the target QSOs.\footnote{ If there is a galaxy over-density around the QSOs, a signature might appear as associated absorption lines in the spectrum of background QSOs \citep[e.g.][]{1982ApJS...48..455Y, 2001ApJ...556..158C, 2007MNRAS.375..735W, 2010arXiv1006.4385C}.  There is one background SDSS QSOs at $z=$\,3.045 behind this structure \citep{2007AJ....133.2222S}.  However, we could not see any clear Ly$\alpha$ absorption-line nor C{\sc iv} absorption between $z=$\,2.216--2.245 in its spectrum.}\\

\noindent{\bf 0200+015:}  There is a local over-density of emitters in the vicinity of the HzRG, with a 3.0\,$\sigma$ deviation from the average density.  The local emitter density of this peak is 3.3 times higher than the average density.  This is one of the highest density peaks in the 0200+015 and the surrounding control fields (see Figure~B2).  We confirmed that all of the five emitter candidates in this over-density satisfy the $BzK$ criteria indicating that they are highly likely to be HAEs at $z=$\,2.23.  Four out of the five HAEs in the over-density, including the HzRG, are also observed with our independent MOIRCS data and selected as narrow-band excess sources, supporting the reliability of our sample selection (see Figure~5 and Figure~9).  Finally,  we obtained optical spectroscopy of the 0200+015-C3-HAE2 using LRIS on Keck, the second brightest HAE in the 0200+015 field.  The 0200+015-C3-HAE2 is a member of the over-density and 1.8\,arcmin (900\,kpc at $z=$\,2.23 in projection) away from the HzRG.  The spectrum confirms that this galaxy is at $z=$\,2.235.  We present the spectrum of this source in Figure~6, which shows strong Ly$\alpha$, C{\sc iv}\,1549 and He{\sc ii}\,1640 emissions lines with velocity widths of FWHM $\sim $\,1000\,km\,s$^{-1}$, suggesting that this HAE is in fact a narrow-line AGN (NLAGN) and further confirms that our emitter selection works well to identify H$\alpha$ emitter at $z=$\,2.23.\\

\noindent{\bf SSA\,13:}  We also find a hint for a local over-density of emitters in the SSA\,13 field.  The local over-density in the vicinity of the SMG/OFRG concentration and has a 2.9\,$\sigma$ deviation from the average density.  The local emitter density of this peak is 2.7 times higher than the average density.  This is the highest density peak in the SSA\,13 and  surrounding control fields (see Figure~B3).  All the SMGs and OFRGs are detected in $K$-band and their $K$-band magnitudes are consistent with the previous results from \citet{2004ApJ...616...71S}.  However, none of them are selected as emitter candidates, although we detect four out of the five SMGs or OFRGs in our H$_2$S1 image (see Figure~2).  This may be due to the relatively shallow depth of the H$_2$S1 image in SSA\,13 (Table~2).  For two of the OFRGs in this field the observed H$\alpha$ equivalent widths have been spectroscopically measured to be $EW_{\rm obs} =$\,65\AA\ and 80\AA\ \citep{2004ApJ...617...64S}.  However, a source with $K \ga $\,19 needs to exhibit a narrow-band excess of $K-{\rm H}_2{\rm S1} \ga $\,1 or EW$_{\rm obs}\ge $\,400\AA\ to comply with our emitter selection criteria.  The H$\alpha$ of the undetected SMG at $z=$\,2.247 falls near the edge of the H$_2$S1 transmission curve (see Figure~1).\\

%
%
\begin{figure}
\centering
  \includegraphics[scale=.7]{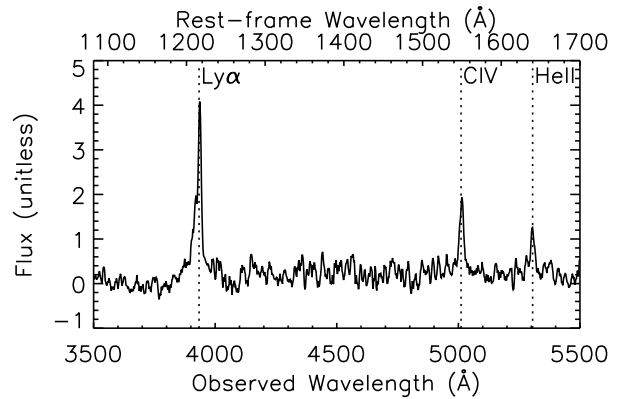}
  \caption{The one dimensional optical spectrum of the second brightest HAE (0200+015-C3-HAE2) in the 0200+015 field. We see strong emission lines corresponding to Ly$\alpha$, C{\sc iv}\,1549 and He{\sc ii}\,1640  which yield a redshift of  $z=$\,2.235, confirming this as an H$\alpha$-selected source.  The strength of He{\sc ii} and the velocity widths of FWHM\,$\sim$\,1000\,km\,s$^{-1}$ for these lines indicate that this HAE is a narrow-line AGN.  This spectroscopic result confirms that our emitter selection works well to identify H$\alpha$ emitter at $z=$\,2.23.}
\end{figure}

%
%
\begin{figure*}
\centering
  \includegraphics[scale=0.95]{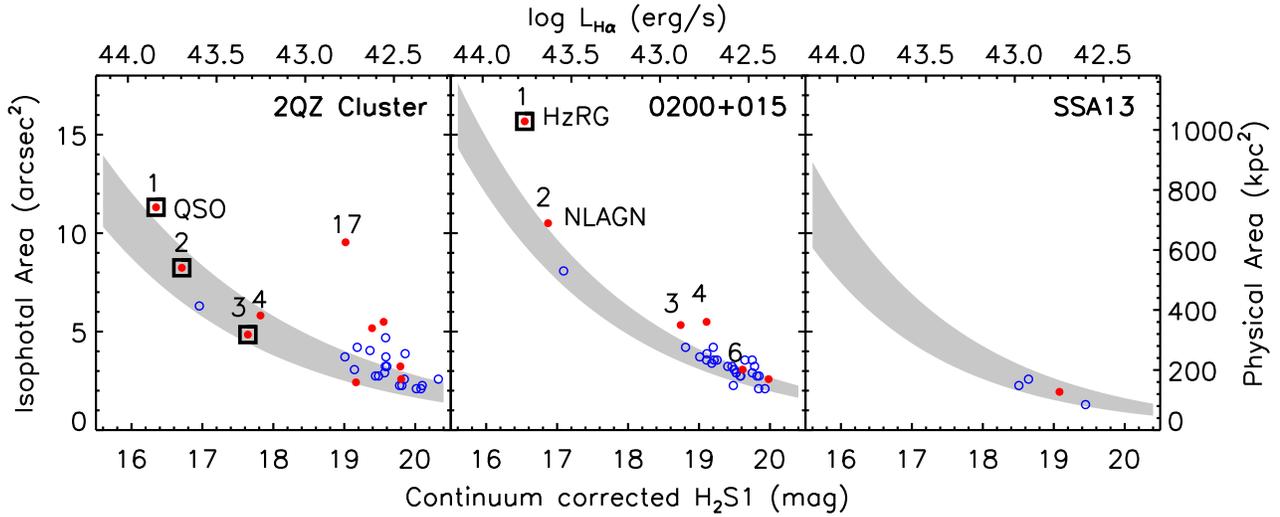}
  \caption{The continuum corrected H$_2$S1 magnitude versus the continuum corrected H$_2$S1 isophotal area for emitter candidates with $\Sigma\ge$\,4.  The grey-scale bands represent the point source track {\it without} narrow-band excess, while the filled circles indicate those emitters lying on the same chip as the target in each field and the open circles are the emitters in the surrounding control fields.  In addition, we highlight the target QSOs and HzRG and  label other sources using the naming scheme from Tables A1 and A2.  The isophotal areas of the point sources show a characteristic variation with magnitude and so we can conclude those narrow-band excess sources significantly above this envelope are well resolved. In particular, HAE17 in the 2QZ cluster field, shows an isophotal area some three times larger than the locus of point sources in the same magnitude range indicating it is very extended.  The magnitudes and area are measured with isophotes determined with $\sim $\,2\,$\sigma$ surface brightness thresholds of 1.0, 1.4 and 2.3\,$\times$\,10$^{-17}$\,erg\,s$^{-1}$ cm$^{-2}$\,arcsec$^{-2}$ for the 2QZ cluster, 0200+015 and SSA\,13 fields respectively.}

\end{figure*}

%
%
\begin{figure}
\centering
  \includegraphics[scale=.4]{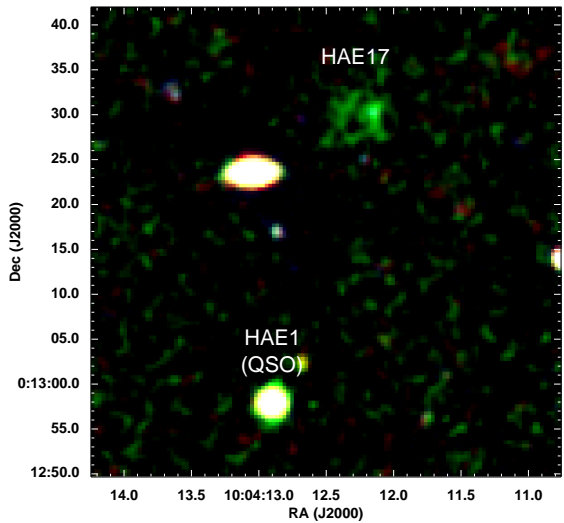}
  \caption{False colour image (blue for SCAM $z'$, green for H$_2$S1, and red for $K$) of the brightest QSO, HAE1 (2QZ\,J100412.8+001257) and the extended HAE candidate, HAE17, in the 2QZ cluster field.  The angular separation between the HAE1 and HAE17 is 34\,arcsec ($\sim$\,300\,kpc in projection).  HAE17 appears to have an extended emission-line nebula with a spatial extent of $\sim $\,7.5\,arcsec ($\sim$\,60\,kpc at $z=$\,2.23) and an H$\alpha$ luminosity of 6\,$ \times $\,10$^{42}$\,erg\,s$^{-1}$.}
\end{figure}

%
%
\begin{figure}
\centering
  \includegraphics[scale=.43]{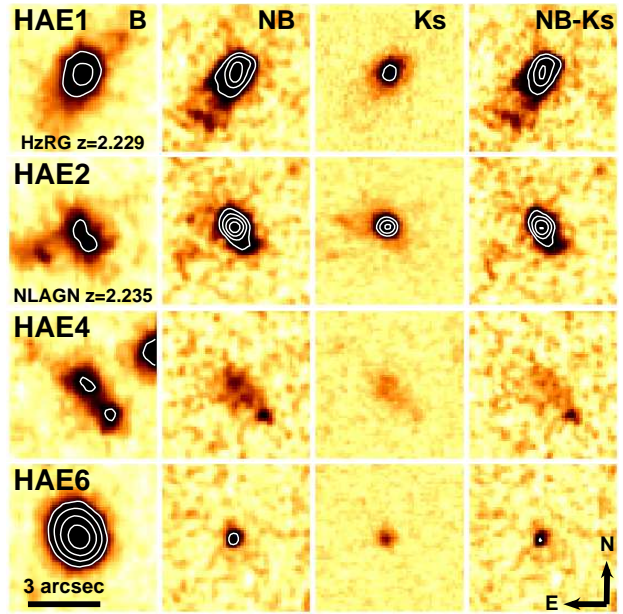}
  \caption{Subaru/MOIRCS thumbnail images of the HAEs in the over-density around the HzRG in the 0200+015 field. The size of the images is 5\,$\times $\,5\,arcsec ($\sim $\,40\,$ \times $\,40\,kpc).  The seeing size of the images are FWHM\,=\,0.5\,arcsec.  All the four emitter satisfy the $BzK$ criteria, indicating that HAE4 and HAE6 are also likely to be at $z\sim$\,2.23.  We see that HAE1, 2, and 4 exhibit extended H$\alpha$ emission on scales of $\sim $\,3--5\,arcsec (or 25--40\,kpc), while HAE6 appears to be compact.  The same magnitude range is used to display for the MOIRCS images.  The  contours show the bright peaks.}
\end{figure}

%
%
\begin{figure}
\centering
  \includegraphics[scale=.6]{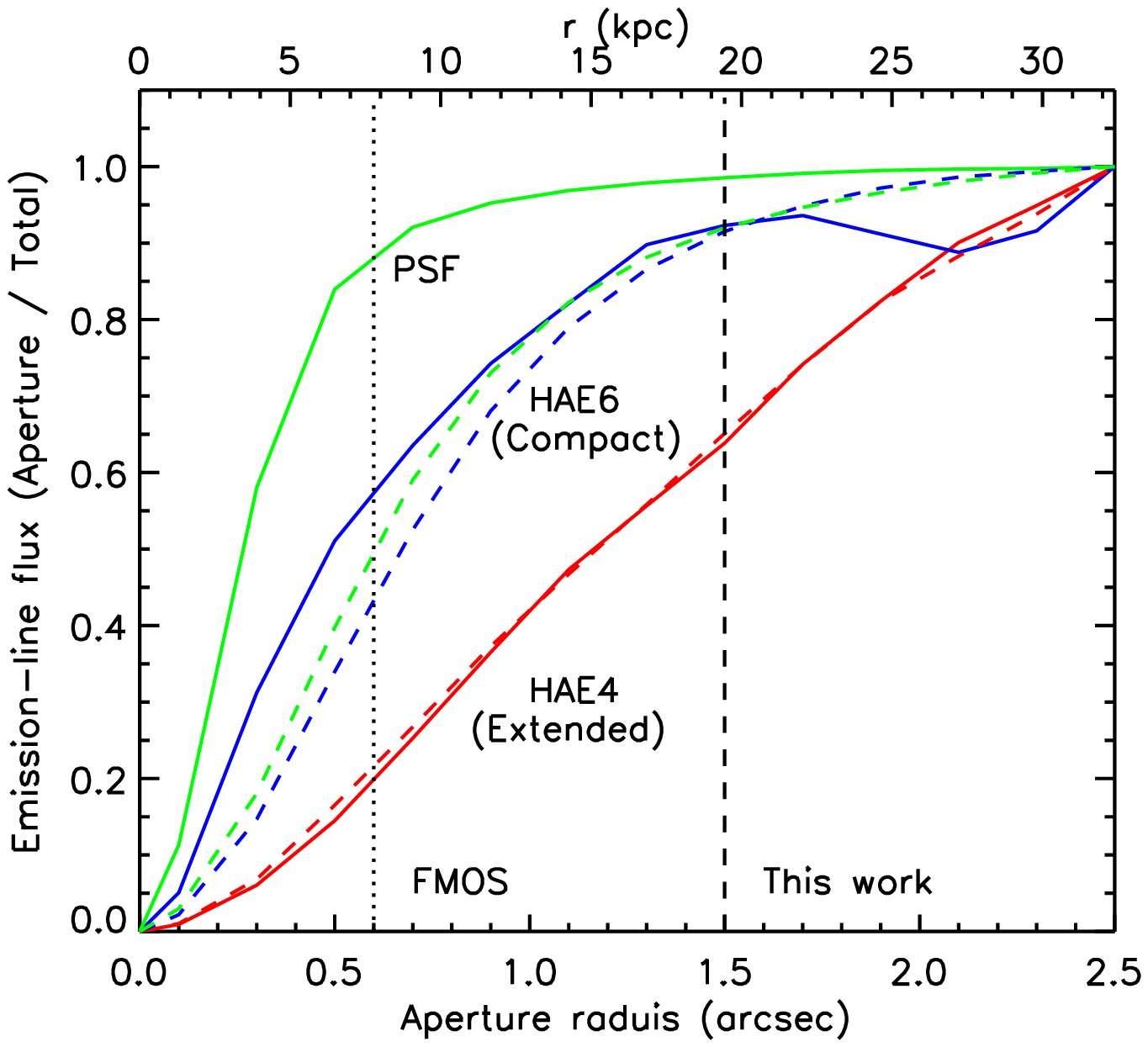}
  \caption{Growth curves of the emission-line fluxes of two typical extended and compact HAEs with MOIRCS imaging in the 0200+015 field.  The solid curves show fractions of covered fluxes of two representative HAEs: HAE4 (Extended), HAE6 (Compact), and a point spread function (PSF) corresponding to the 0.5\,arcsec seeing MOIRCS H$\alpha$ images.  The dashed curves show the fractions of the rest-frame UV continuum in the 1.0\,arcsec SCAM $B$-band images.  The total fluxes are measured with a 2.5\,arcsec diameter and normalized to unity.  The dashed vertical line indicates the 1.5\,arcsec radius aperture photometry used in this work.  The dotted vertical line shows the 0.6\,arcsec radius of fibre of Subaru/FMOS \citep{2010PASJ...62.1135K}.  This suggests that the FMOS fibre may miss up to $\sim $\,40--80\% of the total emission-line fluxes from most extended H$\alpha$ emitters at $z\sim $\,2.  However, aperture corrections based on the rest-frame UV continuum images should be able to recover most of the fluxes within 20\% uncertainty levels.}
\end{figure}

We examine the emission-line morphology of the emitter candidates with $\Sigma\ge$\,4 (Figure~7).  The magnitudes and isophotal areas are measured in the continuum corrected H$_2$S1 images with isophotes determined with $\sim $\,2\,$\sigma$ surface brightness thresholds of 1.0, 1.4 and 2.3\,$ \times $\,10$^{-17}$\,erg\,s$^{-1}$\,cm$^{-2}$\,arcsec$^{-2}$ for the 2QZ cluster,  0200+015  and  SSA\,13 fields respectively.  While the isophotal areas of most of the emitter candidates are similar to, or somewhat larger than, the sequence of the point sources, several emitters in the 2QZ cluster and 0200+015 fields show  significantly larger isophotal areas than expected for point sources.  

The giant HAE candidate in the 2QZ field is HAE17, referred to as 2QZC-C1-HAE17 in Table~A1, and this has a spatial extent of $\sim 7.5$\,arcsec (60\,kpc at $z=2.23$) in our narrow-band image.   This HAE is only 34\,arcsec ($\sim 300$\,kpc)  from the brightest target QSO (HAE1) in this region (see Figure~8) and the continuum source associated with HAE17 satisfies the $BzK$ criteria, suggesting that this is likely to be an H$\alpha$ emission-line nebula at $z=$\,2.23.  The emission-line nebula has a continuum-corrected H$_2$S1 magnitude of 19.0 mag in an isophotal aperture with an area of 9.4\,arcsec$^2$, corresponding to an emission-line flux of 1.7\,$ \times$\,10$^{-16}$\,erg\,s$^{-1}$ cm$^{-2}$ or an H$\alpha$ luminosity of 6\,$ \times $\,10$^{42}$\,erg\,s$^{-1}$.  There is no evidence for an optically bright AGN in HAE17 as the continuum source is spatially resolved in the $B$ and $z'$-band images  and in addition the HAE does not have any bright radio source with $f_{1.4 GHz}\ge 1$\,mJy in the VLA FIRST catalog \citep{1995ApJ...450..559B}, suggesting that it is not similar to the extended emission-line nebulae often seen around powerful radio galaxies  \citep[e.g.][]{1993ARA&A..31..639M}, such as MRC\,0200+015 (Figure~9).  Unfortunately there is no deep X-ray data in this field necessary to further constrain the presence of an obscured AGN in this nebula.

As shown by Figure~7 there are several potentially resolved emitters in the 0200+015 field: the HzRG and the NLAGN, 0200+015-C3-HAE2, as well as  0200+015-C3-HAE4. All show evidence of having larger isophotal areas than expected for point sources with their narrow-band magnitudes.  To investigate the morphologies of these HAEs we can make use of the high-quality imaging from MOIRCS in this field.  In Figure~9 we show thumbnails of the three resolved HAEs from MOIRCS, as well as a fourth more compact HAE which falls in the image: 0200+015-C3-HAE6.  This figure clearly demonstrates the presence of bright H$\alpha$ emission-line nebulae around the HzRG and the NLAGN, 0200+015-C3-HAE2, with spatial extents of 4--5\,arcsec corresponding to $\sim 30$--40\,kpc at $z=$\,2.23, while 0200+015-C3-HAE4 also exhibits fainter but similarly extended emission-line nebulosity.  The $K$-band morphology of  HAE4  also seems extended on 3\,arcsec or $\sim$\,30\,kpc  scales, with the H$\alpha$ emission-line peaking on the outer edge of the $K$-band structure.  These properties of the HAE4 may resemble to galaxies with large disks at $z=1.4-3$ \citep{2003ApJ...591L..95L}.  These extended HAEs, along with HAE17 in the 2QZ field, are obviously relatively common and indicate that star formation is occuring over large regions, comparable to the size of massive galaxies at the present day, even at $z=$\,2.23.  Current surveys of the HAEs in lower density regions at this epoch have turned up few such examples (e.g.\ Figure~7 and G08) and so it is possible that these extended emission-line nebulae are an environmental signature, as has been found for giant Ly$\alpha$ nebulae \citep[Ly$\alpha$ blobs, e.g.][]{2000ApJ...532..170S, 2004ApJ...602..545P, 2004AJ....128..569M, 2009MNRAS.400L..66M, 2010ApJ...719.1654Y}.

We can also use the MOIRCS imaging to determine the efficiency of fibre spectrograph surveys of HAEs using instruments such as FMOS on Subaru  \citep{2010PASJ...62.1135K}.   In Figure~10, we therefore show the growth curves of the emission-line fluxes from  two examples of extended and compact HAEs (0200+015-C3-HAE4 and HAE6) as a function of photometric aperture using the MOIRCS continuum corrected H$_2$S1 image.  For comparison, we also plot a point spread function (PSF) using a bright star in the MOIRCS H$_2$S1 image.  This result suggests that the $\sim$\,10--40\% of the total fluxes are missing in the 3\,arcsec diameter aperture used in this work, while $\sim$\,40--80\% could be missed if we measured the flux in a 1.2\,arcsec diameter aperture fibre similar to Subaru/FMOS.  This suggests that care must be taken to assess the fibre losses for such spectroscopic surveys, even of high-redshift galaxies, as a fraction of the targets may exhibit highly extended emission.   However, as Figure~10 demonstrates, in these cases aperture corrections using rest-frame UV continuum images could be employed to correct these losses and should all the recovery of $\sim $\,80--100\% of the total fluxes.

%
%
\begin{figure*}
\centering
  \includegraphics[scale=1.0]{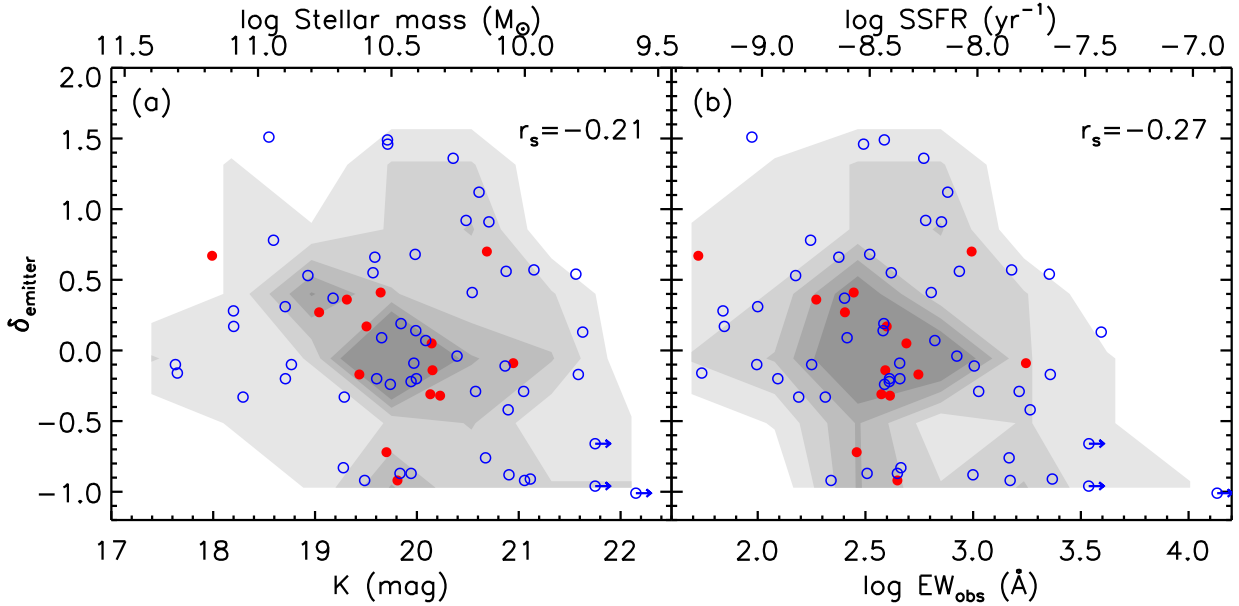}
  \caption{The local over-density versus the observed $K$-band magnitude (a) and equivalent widths (b) of emitter candidates with H$\alpha$ luminosity of L$_{\rm H\alpha} = $\,0.35--2\,$ \times$\,10$^{43}$\,erg\,s$^{-1}$.  Both of the $K$-band magnitudes and equivalent widths show weak correlations with the local over-density.  At $z=$\,2.23, star-forming galaxies in over-dense environments seem to have brighter $K$-band magnitudes, and thus possibly masses, compared to those in under-dense environments.  The filled circles indicate those emitters lying on the same chip as the target in each field, while the open circles are the emitters in the surrounding control fields.  The shaded regions highlight the number densities of the distributions of emitters.  We label each panel with  Spearman's rank correlation coefficient for the distribution,  $r_{\rm s}$, indicating the presence of weak correlations in both these plots.}
\end{figure*}

Finally, in Figure~11 we plot the observed $K$-band magnitudes and equivalent widths of the emitters as a function of the emitter local over-densities (the local densities around each emitters are calculated excluding the source itself).  In order to select an homogeneous emitter sample from the three fields, given their different depths, we exclude emitter candidates in the 2QZ cluster and 0200+015 fields below the detection limit in the SSA\,13 field  ($f < $\,1\,$ \times $\,10$^{-16}$\,erg\,s$^{-1}$\,cm$^{-2}$).  From this combined sample, we also exclude potential AGNs with emission-line fluxes similar to those of target QSOs and HzRG ($f > 7 \times 10^{-16}$ erg s$^{-1}$ cm$^{-2}$), this then means we can crudely relate the $K$-band magnitude to a measure of the stellar mass of the galaxies, assuming that all the HAEs have roughly comparable mass-to-light ratios.  We therefore use the  relation estimated from G08, $\log_{10} $\,M$_{*} \sim -$0.4\,$ \times K({\rm mag}) +$\,18.45, to convert our $K$-band magnitudes into stellar masses.  In a similar manner, as the H$\alpha$ luminosity is a good indicator of star-formation rate \citep{1998ARA&A..36..189K}, we can use our H$\alpha$ equivalent width for the HAEs to estimate their specific star formation rate (SSFR). We show both of these inferred properties in Figure~11.   We see that both the $K$-band magnitudes and equivalent widths (or stellar mass and SSFR) of the resulting sample show weak correlations with the local over-density.  The Spearman's rank correlation coefficients are $r_{\rm s}=-$0.21 ($K$-band magnitude), and $-$0.27 (equivalent width), respectively, indicating that the sample distributions are consistent with  random distributions at 8\% ($K$-band magnitude) and 3\% (equivalent width) confidence levels.  Thus our observations suggest there is weak evidence that the emitters in over-dense region tend to have smaller SSFR compared to those in under-dense environments, potentially indicating accelerated evolution in the build-up of galaxies in dense regions at $z=$\,2.23 \citep[cf.][]{2005ApJ...626...44S, 2010A&A...518A..18T, 2011arXiv1103.4364H}.

\section[]{Discussion and Conclusions}

We have undertaken  a narrow-band imaging survey of three target fields which contain potential signpost of rare over-dense regions with space densities
of $\sim $\,10$^{-7}$--10$^{-8}$\,Mpc$^{-3}$ and local control fields to assess the significance of any over-densities.  Our survey  detects 137 emitter candidates 
in a volume of 2.1\,$ \times $\,10$^{5}$\,comoving\,Mpc$^3$ at $z=$\,2.23 down to limiting H$\alpha$ fluxes of $\sim $\,0.5--1\,$ \times 10^{-16}$\,erg\,s$^{-1}$\,cm$^{-2}$ (equivalent to attenuation un-corrected SFR of $\ga $\,14--28\,M$_\odot$\,yr$^{-1}$ or a dust-corrected SFR of $\ga $\,70--190\,M$_\odot$\,yr$^{-1}$).  This is one of the largest HAE samples currently available at $z\ga$\,2.  Based on supporting optical imaging of these fields, we estimate that at least 80\% of these narrow-band excess sources likely correspond to HAEs at $z=$\,2.23.  We have confirmed the reliability of our sample selection in one field, 0200+015, by using independent H$_2$S1 observations from MOIRCS/Subaru, recovering all four HAEs selected in the overlap area from our WFCAM imaging, and also spectroscopically confirmed that one of the bright emitters in this field is an H$\alpha$ source at $z=$\,2.235.

Wide-field narrow-band H$\alpha$ survey is one of the most effective routes to map over-dense regions of star-forming galaxies at high redshift.  Our analysis of the density distribution of  emitters in our survey, using statistical corrections for the blank-field density (including any potential foreground or background contamination) derived from the parallel control fields, indicates the presence of 3\,$\sigma$ local over-densities in all three of our target fields.  In the 2QZ cluster field the over-dense regions of emitters displays a filamentary structure connecting the four target $z=2.23$ QSOs.  This is similar to the large-scale filamentary structure of Ly$\alpha$ emitters found around the SSA22 $z=3.1$ protocluster \citep{1998ApJ...492..428S, 2004AJ....128.2073H, 2005ApJ...634L.125M}.  Although the over-densities in the vicinities of the QSOs are small ($\delta \sim $\,0--1), there is the density peak with $\delta \sim $\,3 surrounded by these QSOs.  These properties may be similar to the large-scale structure found around another QSO cluster at $z=$\,1.1 \citep{2000ApJ...528..123T, 2001ApJ...547..521T}.  In the HzRG and SMG/OFRG fields, there are smaller-scale over-dense regions in the vicinities of the targets,with peak over-densities of  $\delta \sim $\,2.  These structures are comparable, if slightly smaller, than those found in other H$\alpha$ surveys for proto-clusters around HzRGs at $z\ga$\,2 \citep[e.g.][]{2004A&A...428..793K, 2010arXiv1012.1869T, 2011arXiv1103.4364H}.  In part this may reflect the fact that our  survey is $\sim $\,5--10 times wider but $\sim$\,5--10 times shallower than these surveys.  It is therefore possible that the structures found in this survey are just the tips of more significant, underlying proto-clusters.  Future deeper H$_2$S1 imaging observations will  more accurately reveal the true over-densities of these structures.

Narrow-band H$\alpha$ survey is also useful to understand the relation between the growth of galaxies and their surrounding large-scale structures.  We could not find any clear difference between the H$\alpha$ luminosity functions in the target and control (or blank) fields.  However, analysing our sample of emitters and their environments, we find that their $K$-band magnitudes and equivalent widths (crudely corresponding to stellar mass and specific star-formation rate) show weak correlations with the local over-densities of the emitters.  These results hint at the possibility that the emitters in over-dense region may be more evolved systems compared to those in under-dense regions.  This would be consistent with the picture that star-formation activity is accelerated (or enhanced) in galaxies in over-dense regions at $z \ga $\,2.  Similar results have also been reported from the previous studies of proto-clusters at $z\ga 2$ \citep{2005ApJ...626...44S, 2010A&A...518A..18T, 2011arXiv1103.4364H}.

Deep narrow-band imaging also enables us to examine the H$\alpha$ morphology of emitters, which provides us with information on star-formation activity in high-redshift galaxies during their formation phase and/or on their surrounding circum galactic medium.  In the 0200+015 field, our deeper, higher-resolution MOIRCS narrow-band imaging confirms that the H$\alpha$ emission from three HAEs (including both the HzRG and the newly discovered narrow-line AGN) is significantly extended on scales of 25--40\,kpc.   We find several examples of such extended HAEs in our target fields, including a striking example in the 2QZ field, suggesting that these are relatively common in high-density regions, with no clear examples in our control field or reported in the field survey of G08.  The spatial extent of these systems, $\sim $\,30--60\,kpc, suggests that star formation at $z=$\,2.23 is occuring over regions in these galaxies comparable to the size of the largest galaxies at the present day or large-scale gas outflows are interacting with the surrounding circum galactic medium in overdense environments.  Looking at the H$\alpha$ luminosity of the most extended example, 6\,$ \times $\,10$^{42}$\,erg\,s$^{-1}$ for HAE17 in the 2QZ field, if we assume Ly$\alpha$/H$\alpha = 8.75$ \citep[case B and no dust extinction, e.g.][]{1992ApJ...387L..29M}, then its Ly$\alpha$ luminosity is expected to be 5\,$ \times $\,10$^{43}$\,erg\,s$^{-1}$, which would make its spatial extent and Ly$\alpha$ luminosity comparable to those of giant Ly$\alpha$ blobs \citep{2011MNRAS.410L..13M}.  However, we need spectroscopic follow-up or deeper H$\alpha$ imaging to confirm that this extended emission-line nebula is real and is the first H$\alpha$ blob.  More generally, future higher resolution H$\alpha$ imaging and integral field spectroscopic observations are essential to investigate the star-formation activiy and gas dynamics and metallicity of the inter-stellar medium in these spatially-extended emitters to constrain their role in galaxy formation.    

\section*{Acknowledgments}

We thank Alastair Edge, Masayuki Akiyama, Tomoki Hayashino and Scott Chapman for useful discussions.  We also thank Jae-Woo Kim for providing their UKIDSS/DXS catalog.  YM and IRS acknowledge support from STFC.  JEG is supported by NSERC.  Some of the data reported here were obtained as part of the UKIRT Service Programme.  UKIRT is funded by the STFC.  The W.\,M.\ Keck Observatory was made possible by the generous financial support of the W.\,M.\ Keck Foundation.  The authors wish to recognize and acknowledge the very significant cultural role and reverence that the summit of Mauna Kea has always had within the indigenous Hawaiian community.  We are most fortunate to have the opportunity to conduct observations from this mountain.

\appendix

\section[]{Tables and thumbnail images of H$\alpha$ emitter candidates}

%
%
\begin{table*}
\centering
\begin{minipage}{175mm}
 \caption{Properties of the emitter candidates in the 2QZ cluster (C1) and the control fields (C2, C3, and C4)}
 \begin{tabular}{@{}lcccccccccccccccc}
  \hline
ID & Coordinate (J2000) & ${\rm H}_2{\rm S1}$ & $K$ & $\Sigma$ & $EW_{\rm obs}$ & log $F_{\rm {H}\alpha}^a$ &log $L_{\rm {H}\alpha}^b$ & SFR$_{\rm {H}\alpha}^c$ & M$_*^d$ & Note \\
 & (h:m:s) (d:m:s) & (mag) & (mag) &  & (\AA) & (cgs) & (cgs) &  & \\
\hline
2QZC-C1-HAE1 & 10:04:12.90 +00:12:57.9 & 15.58 & 16.29 & 110.0 & 219 & -14.68 & 43.88 & 7289 & 11.9 & QSO \\
2QZC-C1-HAE2 & 10:03:39.79 +00:21:10.8 & 16.02 & 16.80 & 79.1 & 255 & -14.82 & 43.74 & 4868 & 11.7 & QSO \\
2QZC-C1-HAE3 & 10:03:51.58 +00:15:02.1 & 17.03 & 17.95 & 34.5 & 327 & -15.18 & 43.38 & 1761 & 11.3 & QSO \\
2QZC-C1-HAE4 & 10:03:38.27 +00:18:23.8 & 17.11 & 17.89 & 28.8 & 255 & -15.26 & 43.30 & 1414 & 11.3 & ... \\
2QZC-C1-HAE5 & 10:03:25.05 +00:09:20.5 & 18.66 & 18.93 & 3.0 & 66 & -16.24 & 42.33 & 90 & 10.9 & ... \\
2QZC-C1-HAE6 & 10:04:09.08 +00:15:45.9 & 18.80 & 19.12 & 3.0 & 78 & -16.24 & 42.32 & 88 & 10.8 & BzK \\
2QZC-C1-HAE7 & 10:03:38.19 +00:14:35.2 & 18.97 & 20.95 & 8.5 & 1758 & -15.79 & 42.77 & 316 & 10.1 & BzK \\
2QZC-C1-HAE8 & 10:03:26.43 +00:13:01.2 & 19.13 & 20.15 & 5.3 & 391 & -15.99 & 42.57 & 179 & 10.4 & BzK \\
2QZC-C1-HAE9 & 10:03:32.97 +00:10:04.5 & 19.27 & 20.03 & 3.8 & 241 & -16.13 & 42.43 & 120 & 10.4 & BzK \\
2QZC-C1-HAE10 & 10:04:04.29 +00:14:14.4 & 19.34 & 19.90 & 2.9 & 156 & -16.26 & 42.30 & 84 & 10.5 & BzK \\
2QZC-C1-HAE11 & 10:03:33.09 +00:21:28.4 & 19.50 & 20.98 & 4.6 & 808 & -16.05 & 42.51 & 151 & 10.1 & BzK \\
2QZC-C1-HAE12 & 10:03:33.57 +00:21:22.8 & 19.52 & 20.50 & 3.7 & 369 & -16.16 & 42.41 & 113 & 10.2 & BzK \\
2QZC-C1-HAE13 & 10:03:24.48 +00:10:52.4 & 19.52 & 20.40 & 3.4 & 307 & -16.19 & 42.38 & 103 & 10.3 & BzK \\
2QZC-C1-HAE14 & 10:03:43.28 +00:11:44.5 & 19.53 & 20.48 & 3.5 & 344 & -16.17 & 42.39 & 108 & 10.3 & BzK \\
2QZC-C1-HAE15 & 10:03:43.84 +00:14:20.2 & 19.53 & 20.24 & 2.9 & 221 & -16.26 & 42.31 & 85 & 10.4 & BzK \\
2QZC-C1-HAE16 & 10:03:39.78 +00:11:15.8 & 19.57 & 20.36 & 3.0 & 257 & -16.24 & 42.32 & 88 & 10.3 & BzK \\
2QZC-C1-HAE17 & 10:04:12.16 +00:13:29.7 & 19.60 & 21.95 & 5.0 & 3460 & -16.02 & 42.54 & 165 & 9.7 & BzK \\
2QZC-C1-HAE18 & 10:03:27.82 +00:10:58.6 & 19.67 & 21.04 & 3.8 & 682 & -16.14 & 42.42 & 118 & 10.0 & BzK \\
2QZC-C1-HAE19 & 10:03:24.89 +00:08:31.8 & 19.71 & $>$22.15 & $>$4.6 & $>$4237 & -16.06 & 42.50 & 148 & $<$9.6 & ... \\
2QZC-C1-HAE20 & 10:04:16.74 +00:18:33.0 & 19.82 & 21.25 & 3.4 & 745 & -16.19 & 42.37 & 102 & 9.9 & ... \\
2QZC-C1-HAE21 & 10:04:02.90 +00:18:52.5 & 19.85 & $>$22.15 & $>$4.0 & $>$3167 & -16.12 & 42.44 & 124 & $<$9.6 & ... \\
2QZC-C1-HAE22 & 10:03:40.59 +00:10:07.1 & 19.86 & 20.83 & 2.6 & 356 & -16.30 & 42.26 & 75 & 10.1 & ... \\
\hline
 \end{tabular}
$^a$ The H$\alpha$ flux (erg s$^{-1}$ cm$^{-2}$) corrected for 33\% [N{\sc ii}] contribution to the measured flux.\\
$^b$ The H$\alpha$ luminosity (erg s$^{-1}$).\\
$^c$ The attenuation corrected star-formation rate derived from the H$\alpha$ luminosity (M$_{\sun}$ yr$^{-1}$).\\
$^d$ The stellar mass estimated from the $K$-band magnitude by assuming a constant mass-to-light ratio from \citet{2008MNRAS.388.1473G}.\\
\end{minipage}
\end{table*}

\setcounter{table}{0} 

\begin{table*}
\centering
\begin{minipage}{175mm}
 \caption{continued}
 \begin{tabular}{@{}lcccccccccccccccc}
  \hline
ID & Coordinate (J2000) & ${\rm H}_2{\rm S1}$ & $K$ & $\Sigma$ & $EW_{\rm obs}$ & log $F_{\rm {H}\alpha}^a$ &log $L_{\rm {H}\alpha}^b$ & SFR$_{\rm {H}\alpha}^c$ & M$_*^d$ & Note\\
 & (h:m:s) (d:m:s) & (mag) & (mag) &  & (\AA) & (cgs) & (cgs) &  & \\
\hline
2QZC-C2-HAE1 & 10:05:11.82 +00:14:18.3 & 16.65 & 18.14 & 64.3 & 824 & -14.91 & 43.65 & 3778 & 11.2 & ... \\
2QZC-C2-HAE2 & 10:05:12.69 +00:10:58.7 & 17.91 & 18.20 & 6.3 & 70 & -15.92 & 42.64 & 219 & 11.2 & ... \\
2QZC-C2-HAE3 & 10:05:36.81 +00:14:19.4 & 18.81 & 20.67 & 9.6 & 1469 & -15.73 & 42.83 & 370 & 10.2 & ... \\
2QZC-C2-HAE4 & 10:05:34.17 +00:19:32.4 & 18.92 & 20.58 & 8.3 & 1061 & -15.80 & 42.76 & 309 & 10.2 & ... \\
2QZC-C2-HAE5 & 10:05:15.08 +00:12:01.8 & 19.06 & 19.98 & 5.4 & 331 & -15.99 & 42.57 & 180 & 10.5 & ... \\
2QZC-C2-HAE6 & 10:05:18.63 +00:14:20.9 & 19.25 & 20.14 & 4.3 & 308 & -16.08 & 42.48 & 139 & 10.4 & ... \\
2QZC-C2-HAE7 & 10:05:40.32 +00:11:09.1 & 19.27 & 20.38 & 4.9 & 450 & -16.03 & 42.53 & 161 & 10.3 & ... \\
2QZC-C2-HAE8 & 10:05:24.12 +00:19:39.4 & 19.28 & 20.10 & 4.1 & 276 & -16.11 & 42.45 & 128 & 10.4 & ... \\
2QZC-C2-HAE9 & 10:05:26.89 +00:14:36.4 & 19.38 & 20.02 & 3.1 & 193 & -16.22 & 42.34 & 93 & 10.4 & ... \\
2QZC-C2-HAE10 & 10:05:29.41 +00:19:59.4 & 19.44 & 20.21 & 3.3 & 250 & -16.19 & 42.37 & 101 & 10.4 & ... \\
2QZC-C2-HAE11 & 10:05:15.88 +00:14:08.7 & 19.45 & 20.39 & 3.7 & 337 & -16.15 & 42.42 & 116 & 10.3 & ... \\
2QZC-C2-HAE12 & 10:05:32.07 +00:09:00.4 & 19.45 & 20.00 & 2.5 & 151 & -16.31 & 42.25 & 72 & 10.5 & ... \\
2QZC-C2-HAE13 & 10:05:12.95 +00:09:10.2 & 19.68 & 21.05 & 3.8 & 680 & -16.14 & 42.42 & 118 & 10.0 & ... \\
2QZC-C2-HAE14 & 10:05:14.51 +00:17:36.5 & 19.75 & 21.43 & 3.9 & 1084 & -16.13 & 42.43 & 121 & 9.9 & ... \\
2QZC-C2-HAE15 & 10:05:51.55 +00:13:16.4 & 19.76 & 21.61 & 4.0 & 1436 & -16.12 & 42.45 & 126 & 9.8 & ... \\
2QZC-C2-HAE16 & 10:05:10.98 +00:19:55.6 & 19.87 & 21.66 & 3.6 & 1312 & -16.17 & 42.40 & 110 & 9.8 & ... \\
2QZC-C2-HAE17 & 10:05:18.32 +00:14:21.0 & 19.89 & 21.59 & 3.4 & 1125 & -16.18 & 42.38 & 104 & 9.8 & ... \\
2QZC-C2-HAE18 & 10:05:19.23 +00:15:01.5 & 19.90 & $>$22.15 & $>$3.8 & $>$2862 & -16.14 & 42.42 & 116 & $<$9.6 & ... \\
\hline
2QZC-C3-HAE1 & 10:05:19.28 +00:40:16.0 & 18.40 & 19.18 & 8.8 & 253 & -15.77 & 42.79 & 331 & 10.8 & ... \\
2QZC-C3-HAE2 & 10:05:17.44 +00:37:44.5 & 18.48 & 18.70 & 2.9 & 50 & -16.26 & 42.30 & 84 & 11.0 & ... \\
2QZC-C3-HAE3 & 10:05:31.07 +00:41:00.7 & 18.71 & 18.96 & 2.7 & 60 & -16.29 & 42.27 & 77 & 10.9 & ... \\
2QZC-C3-HAE4 & 10:05:21.53 +00:38:26.3 & 18.82 & 19.71 & 6.5 & 310 & -15.91 & 42.65 & 227 & 10.6 & ... \\
2QZC-C3-HAE5 & 10:05:53.65 +00:43:07.9 & 18.84 & 19.59 & 5.7 & 238 & -15.96 & 42.60 & 194 & 10.6 & ... \\
2QZC-C3-HAE6 & 10:05:56.86 +00:39:31.8 & 19.02 & 19.45 & 3.1 & 112 & -16.22 & 42.34 & 94 & 10.7 & ... \\
2QZC-C3-HAE7 & 10:05:26.30 +00:37:05.0 & 19.07 & 19.92 & 5.0 & 285 & -16.02 & 42.54 & 164 & 10.5 & ... \\
2QZC-C3-HAE8 & 10:06:00.01 +00:42:33.6 & 19.31 & 20.71 & 5.4 & 713 & -15.99 & 42.57 & 181 & 10.2 & ... \\
2QZC-C3-HAE9 & 10:05:36.99 +00:46:20.2 & 19.35 & $>$22.15 & $>$6.6 & $>$13565 & -15.90 & 42.66 & 231 & $<$9.6 & ... \\
2QZC-C3-HAE10 & 10:05:55.27 +00:44:11.7 & 19.43 & 20.45 & 4.0 & 387 & -16.12 & 42.45 & 126 & 10.3 & ... \\
2QZC-C3-HAE11 & 10:06:02.53 +00:43:34.8 & 19.44 & 21.56 & 5.6 & 2257 & -15.97 & 42.59 & 192 & 9.8 & ... \\
2QZC-C3-HAE12 & 10:06:03.90 +00:41:06.3 & 19.46 & 21.59 & 5.5 & 2281 & -15.97 & 42.59 & 188 & 9.8 & ... \\
2QZC-C3-HAE13 & 10:05:11.91 +00:35:31.2 & 19.52 & 20.23 & 2.9 & 221 & -16.25 & 42.31 & 86 & 10.4 & ... \\
2QZC-C3-HAE14 & 10:05:49.59 +00:39:53.2 & 19.70 & 20.99 & 3.6 & 599 & -16.16 & 42.40 & 110 & 10.1 & ... \\
2QZC-C3-HAE15 & 10:05:16.28 +00:41:50.4 & 19.70 & 21.28 & 3.9 & 934 & -16.12 & 42.44 & 124 & 9.9 & ... \\
2QZC-C3-HAE16 & 10:05:33.56 +00:36:13.9 & 19.70 & 20.83 & 3.3 & 463 & -16.20 & 42.36 & 100 & 10.1 & ... \\
2QZC-C3-HAE17 & 10:05:35.09 +00:39:44.7 & 19.72 & 20.82 & 3.2 & 444 & -16.21 & 42.35 & 96 & 10.1 & ... \\
2QZC-C3-HAE18 & 10:05:36.78 +00:39:09.2 & 19.77 & 21.04 & 3.3 & 587 & -16.19 & 42.37 & 101 & 10.0 & ... \\
2QZC-C3-HAE19 & 10:05:22.24 +00:37:38.4 & 19.79 & 21.11 & 3.3 & 634 & -16.19 & 42.37 & 101 & 10.0 & ... \\
2QZC-C3-HAE20 & 10:05:42.34 +00:41:26.0 & 19.87 & 20.84 & 2.6 & 353 & -16.30 & 42.26 & 74 & 10.1 & ... \\
\hline
2QZC-C4-HAE1 & 10:04:01.76 +00:36:59.4 & 18.61 & 19.29 & 6.6 & 206 & -15.90 & 42.66 & 230 & 10.7 & ... \\
2QZC-C4-HAE2 & 10:03:43.23 +00:41:35.6 & 18.93 & 19.30 & 3.0 & 93 & -16.24 & 42.32 & 89 & 10.7 & ... \\
2QZC-C4-HAE3 & 10:03:53.81 +00:36:05.7 & 19.54 & 21.27 & 4.8 & 1185 & -16.04 & 42.52 & 156 & 9.9 & ... \\
2QZC-C4-HAE4 & 10:04:08.44 +00:43:09.8 & 19.66 & 20.95 & 3.7 & 600 & -16.15 & 42.41 & 115 & 10.1 & ... \\
2QZC-C4-HAE5 & 10:03:54.01 +00:35:37.6 & 19.87 & 20.79 & 2.5 & 331 & -16.31 & 42.25 & 72 & 10.1 & ... \\
2QZC-C4-HAE6 & 010:04:08.45 +00:34:58.3 & 19.89 & $>$22.15 & $>$3.8 & $>$2913 & -16.14 & 42.42 & 118 & $<$9.6 & ... \\
\hline
 \end{tabular}
$^a$ The H$\alpha$ flux (erg s$^{-1}$ cm$^{-2}$) corrected for 33\% [N{\sc ii}] contribution to the measured flux.\\
$^b$ The H$\alpha$ luminosity (erg s$^{-1}$).\\
$^c$ The attenuation corrected star-formation rate derived from the H$\alpha$ luminosity (M$_{\sun}$ yr$^{-1}$).\\
$^d$ The stellar mass estimated from the $K$-band magnitude by assuming a constant mass-to-light ratio from \citet{2008MNRAS.388.1473G}.
\end{minipage}
\end{table*}

%
%
\begin{figure*}
\centering
  \includegraphics[scale=.9]{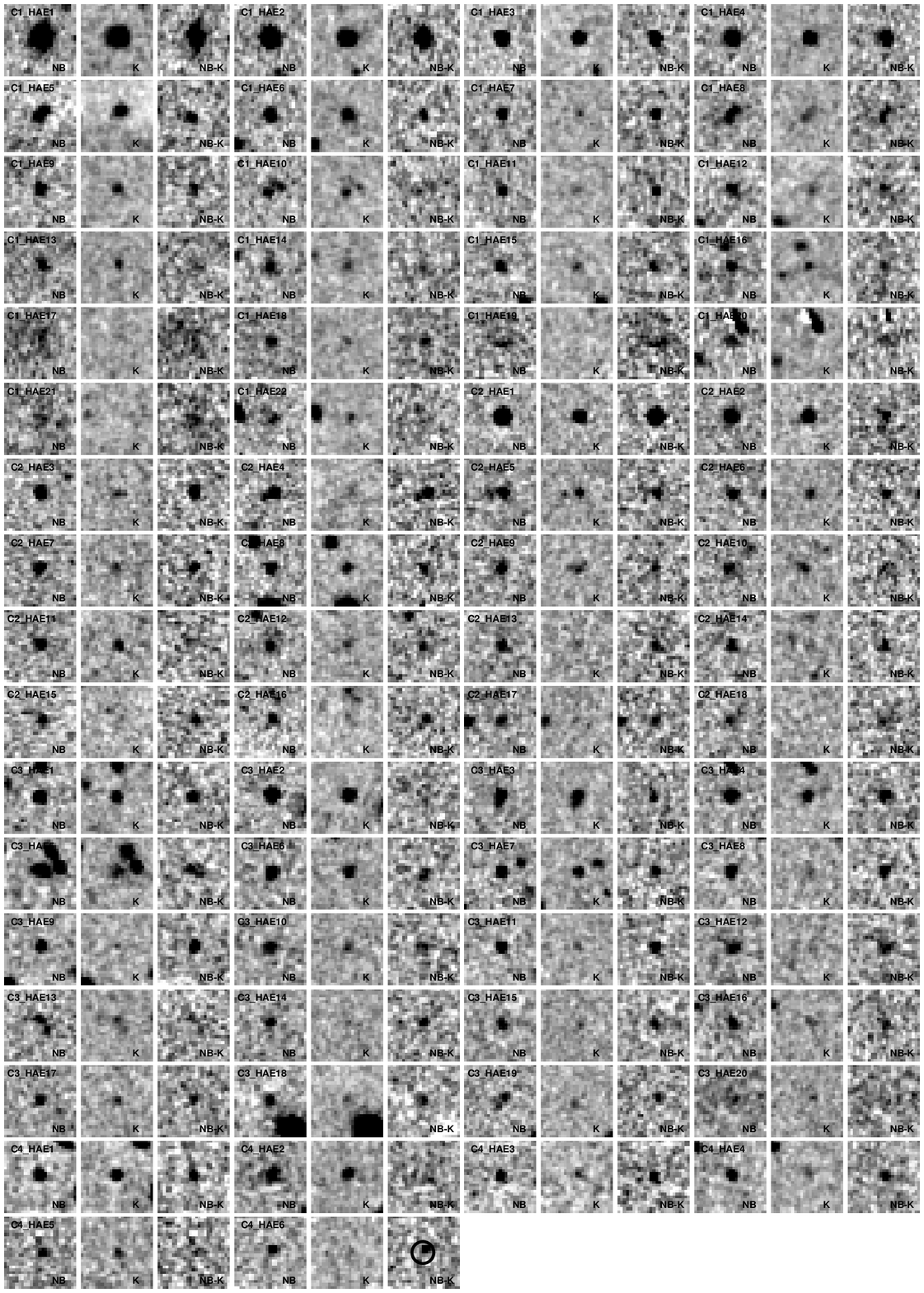}
  \caption{WFCAM thumbnail images of the emitter candidates in the 2QZ cluster and surrounding control fields.  The size of the images is $10 \times 10$ arcsec$^2$ ($\sim 80 \times 80$\,kpc$^2$).  The circle in the bottom-left panel shows the size of the 3\,arcsec aperture for photometry.  The same magnitude range are used to display for all the images.}
\end{figure*}

%
%
\begin{table*}
\centering
\begin{minipage}{170mm}
 \caption{Properties of the emitter candidates in the 0200+015 field (C3) and the control fields (C1, C2, and C4)}
 \begin{tabular}{@{}lcccccccccccccccc}
  \hline
ID & Coordinate (J2000) & ${\rm H}_2{\rm S1}$ & $K$ & $\Sigma$ & $EW_{\rm obs}$ & log $F_{\rm {H}\alpha}^a$ &log $L_{\rm {H}\alpha}^b$ & SFR$_{\rm {H}\alpha}^c$ & M$_*^d$ & Note \\
 & (h:m:s) (d:m:s) & (mag) & (mag) &  & (\AA) & (cgs) & (cgs) &  & \\
\hline
0200-C3-HAE1 & 2:02:42.99 +01:49:10.8 & 16.48 & 18.30 & 55.9 & 1369 & -14.81 & 43.76 & 5080 & 11.1 & HzRG \\
0200-C3-HAE2 & 2:02:50.22 +01:48:53.3 & 16.57 & 18.01 & 46.5 & 760 & -14.89 & 43.68 & 4054 & 11.2 & NLAGN \\
0200-C3-HAE3 & 2:02:37.68 +01:44:33.2 & 18.19 & 19.43 & 9.6 & 557 & -15.57 & 42.99 & 589 & 10.7 & BzK \\
0200-C3-HAE4 & 2:02:37.36 +01:48:22.2 & 18.47 & 19.50 & 6.7 & 397 & -15.73 & 42.84 & 378 & 10.6 & BzK \\
0200-C3-HAE5 & 2:03:01.04 +01:52:09.9 & 18.85 & 19.70 & 4.2 & 288 & -15.93 & 42.63 & 212 & 10.6 & ... \\
0200-C3-HAE6 & 2:02:46.98 +01:47:54.2 & 19.08 & 20.69 & 4.8 & 984 & -15.87 & 42.69 & 254 & 10.2 & BzK \\
0200-C3-HAE7 & 2:02:19.37 +01:55:39.8 & 19.13 & 20.13 & 3.6 & 375 & -16.00 & 42.56 & 175 & 10.4 & BzK \\
0200-C3-HAE8 & 2:02:24.55 +01:56:06.4 & 19.18 & 20.23 & 3.5 & 411 & -16.00 & 42.56 & 174 & 10.4 & ... \\
0200-C3-HAE9 & 2:03:02.81 +01:49:20.5 & 19.35 & 20.15 & 2.6 & 265 & -16.15 & 42.42 & 116 & 10.4 & BzK \\
0200-C3-HAE10 & 2:02:51.92 +01:50:02.9 & 19.36 & 20.19 & 2.6 & 280 & -16.14 & 42.42 & 118 & 10.4 & BzK \\
0200-C3-HAE11 & 2:02:34.33 +01:44:58.9 & 19.47 & 20.57 & 2.8 & 442 & -16.11 & 42.45 & 128 & 10.2 & BzK \\
\hline
0200-C1-HAE1 & 2:01:03.23 +01:20:09.7 & 17.74 & 18.29 & 8.5 & 155 & -15.62 & 42.94 & 508 & 11.1 & ... \\
0200-C1-HAE2 & 2:01:21.56 +01:24:45.0 & 18.32 & 18.58 & 2.6 & 61 & -16.13 & 42.43 & 121 & 11.0 & ... \\
0200-C1-HAE3 & 2:00:36.54 +01:20:18.0 & 18.69 & 19.71 & 5.4 & 387 & -15.82 & 42.74 & 291 & 10.6 & ... \\
0200-C1-HAE4 & 2:01:06.09 +01:23:30.8 & 18.73 & 20.09 & 6.1 & 664 & -15.77 & 42.80 & 339 & 10.4 & ... \\
0200-C1-HAE5 & 2:00:38.81 +01:22:56.7 & 18.94 & 20.00 & 4.4 & 409 & -15.91 & 42.65 & 225 & 10.5 & ... \\
0200-C1-HAE6 & 2:00:36.58 +01:19:53.7 & 19.08 & 20.36 & 4.3 & 590 & -15.92 & 42.65 & 221 & 10.3 & ... \\
0200-C1-HAE7 & 2:00:57.57 +01:19:39.2 & 19.11 & 21.05 & 5.0 & 1635 & -15.85 & 42.71 & 266 & 10.0 & ... \\
0200-C1-HAE8 & 2:00:43.94 +01:26:14.8 & 19.18 & 21.06 & 4.7 & 1485 & -15.88 & 42.68 & 243 & 10.0 & ... \\
0200-C1-HAE9 & 2:00:33.53 +01:21:28.7 & 19.19 & 20.48 & 3.9 & 602 & -15.96 & 42.60 & 196 & 10.3 & ... \\
0200-C1-HAE10 & 2:01:05.81 +01:23:50.3 & 19.23 & 21.63 & 4.9 & 3931 & -15.87 & 42.70 & 255 & 9.8 & ... \\
0200-C1-HAE11 & 2:00:40.04 +01:19:22.0 & 19.35 & 20.88 & 3.7 & 863 & -15.99 & 42.57 & 181 & 10.1 & ... \\
0200-C1-HAE12 & 2:01:07.72 +01:28:22.6 & 19.41 & $>$21.75 & $>$4.1 & $>$3438 & -15.94 & 42.62 & 206 & $<$9.8 & ... \\
\hline
0200-C2-HAE1 & 2:03:08.25 +01:28:18.1 & 16.44 & 17.26 & 37.6 & 273 & -14.98 & 43.58 & 3121 & 11.5 & ... \\
0200-C2-HAE2 & 2:02:16.71 +01:30:06.7 & 17.99 & 18.59 & 7.3 & 176 & -15.69 & 42.87 & 420 & 11.0 & ... \\
0200-C2-HAE3 & 2:02:52.26 +01:29:33.7 & 18.16 & 18.77 & 6.3 & 178 & -15.75 & 42.81 & 349 & 10.9 & ... \\
0200-C2-HAE4 & 2:02:50.60 +01:24:55.1 & 18.51 & 19.57 & 6.6 & 417 & -15.73 & 42.83 & 371 & 10.6 & ... \\
0200-C2-HAE5 & 2:02:52.81 +01:23:00.1 & 18.83 & 19.84 & 4.8 & 385 & -15.87 & 42.69 & 250 & 10.5 & ... \\
0200-C2-HAE6 & 2:02:51.23 +01:30:57.2 & 18.85 & 19.97 & 4.9 & 456 & -15.86 & 42.70 & 260 & 10.5 & ... \\
0200-C2-HAE7 & 2:02:46.26 +01:21:46.2 & 18.89 & 20.90 & 6.2 & 1839 & -15.76 & 42.81 & 347 & 10.1 & ... \\
0200-C2-HAE8 & 2:03:00.80 +01:27:05.4 & 18.93 & 19.45 & 2.7 & 144 & -16.12 & 42.45 & 126 & 10.7 & ... \\
0200-C2-HAE9 & 2:03:05.95 +01:28:22.9 & 19.17 & 20.61 & 4.2 & 762 & -15.93 & 42.64 & 216 & 10.2 & ... \\
0200-C2-HAE10 & 2:02:51.75 +01:24:21.8 & 19.19 & 20.14 & 3.3 & 344 & -16.04 & 42.53 & 158 & 10.4 & ... \\
0200-C2-HAE11 & 2:02:19.56 +01:30:54.0 & 19.21 & 20.54 & 3.9 & 639 & -15.96 & 42.60 & 195 & 10.2 & ... \\
0200-C2-HAE12 & 2:02:19.72 +01:21:32.2 & 19.28 & 20.90 & 4.0 & 998 & -15.95 & 42.61 & 201 & 10.1 & ... \\
0200-C2-HAE13 & 2:02:40.64 +01:30:18.2 & 19.41 & $>$21.75 & $>$4.1 & $>$3445 & -15.94 & 42.62 & 206 & $<$9.8 & ... \\
0200-C2-HAE14 & 2:02:20.53 +01:18:16.1 & 19.44 & 20.45 & 2.7 & 384 & -16.12 & 42.44 & 125 & 10.3 & ... \\
0200-C2-HAE15 & 2:02:16.49 +01:29:17.8 & 19.45 & 20.65 & 3.0 & 521 & -16.08 & 42.48 & 139 & 10.2 & ... \\
0200-C2-HAE16 & 2:03:04.90 +01:29:38.8 & 19.47 & 21.12 & 3.4 & 1046 & -16.02 & 42.54 & 165 & 10.0 & ... \\
\hline
0200-C4-HAE1 & 2:01:04.87 +01:50:43.6 & 17.41 & 17.65 & 5.6 & 55 & -15.80 & 42.76 & 306 & 11.4 & ... \\
0200-C4-HAE2 & 2:00:57.44 +01:50:06.2 & 17.91 & 18.20 & 4.2 & 69 & -15.93 & 42.63 & 214 & 11.2 & ... \\
0200-C4-HAE3 & 2:00:51.42 +01:45:56.2 & 18.18 & 18.55 & 4.2 & 94 & -15.93 & 42.63 & 210 & 11.0 & ... \\
0200-C4-HAE4 & 2:00:53.01 +01:47:18.6 & 18.43 & 18.71 & 2.6 & 68 & -16.13 & 42.43 & 120 & 11.0 & ... \\
0200-C4-HAE5 & 2:01:12.56 +01:45:14.7 & 18.72 & 19.74 & 5.3 & 388 & -15.83 & 42.73 & 283 & 10.6 & ... \\
0200-C4-HAE6 & 2:00:47.83 +01:44:55.2 & 19.02 & 19.56 & 2.6 & 154 & -16.13 & 42.43 & 120 & 10.6 & ... \\
0200-C4-HAE7 & 2:01:07.59 +01:44:34.1 & 19.24 & 20.87 & 4.2 & 1012 & -15.93 & 42.63 & 213 & 10.1 & ... \\
0200-C4-HAE8 & 2:01:05.52 +01:48:03.1 & 19.27 & 21.15 & 4.3 & 1509 & -15.92 & 42.65 & 222 & 10.0 & ... \\
0200-C4-HAE9 & 2:01:05.89 +01:48:13.6 & 19.37 & 20.59 & 3.2 & 537 & -16.04 & 42.52 & 155 & 10.2 & ... \\
0200-C4-HAE10 & 2:00:48.00 +01:50:10.0 & 19.45 & 20.38 & 2.5 & 331 & -16.15 & 42.42 & 116 & 10.3 & ... \\
0200-C4-HAE11 & 2:00:51.62 +01:47:20.2 & 19.46 & 20.77 & 3.1 & 617 & -16.07 & 42.50 & 145 & 10.1 & ... \\
0200-C4-HAE12 & 2:00:55.98 +01:44:36.2 & 19.50 & 20.50 & 2.6 & 378 & -16.14 & 42.42 & 116 & 10.2 & ... \\
\hline
 \end{tabular}
$^a$ The H$\alpha$ flux (erg s$^{-1}$ cm$^{-2}$) corrected for 33\% [N{\sc ii}] contribution to the measured flux.\\
$^b$ The H$\alpha$ luminosity (erg s$^{-1}$).\\
$^c$ The attenuation corrected star-formation rate derived from the H$\alpha$ luminosity (M$_{\sun}$ yr$^{-1}$).\\
$^d$ The stellar mass estimated from the $K$-band magnitude by assuming a constant mass-to-light ratio from \citet{2008MNRAS.388.1473G}.\\
\end{minipage}
\end{table*}

%
%
\begin{figure*}
\centering
  \includegraphics[scale=.9]{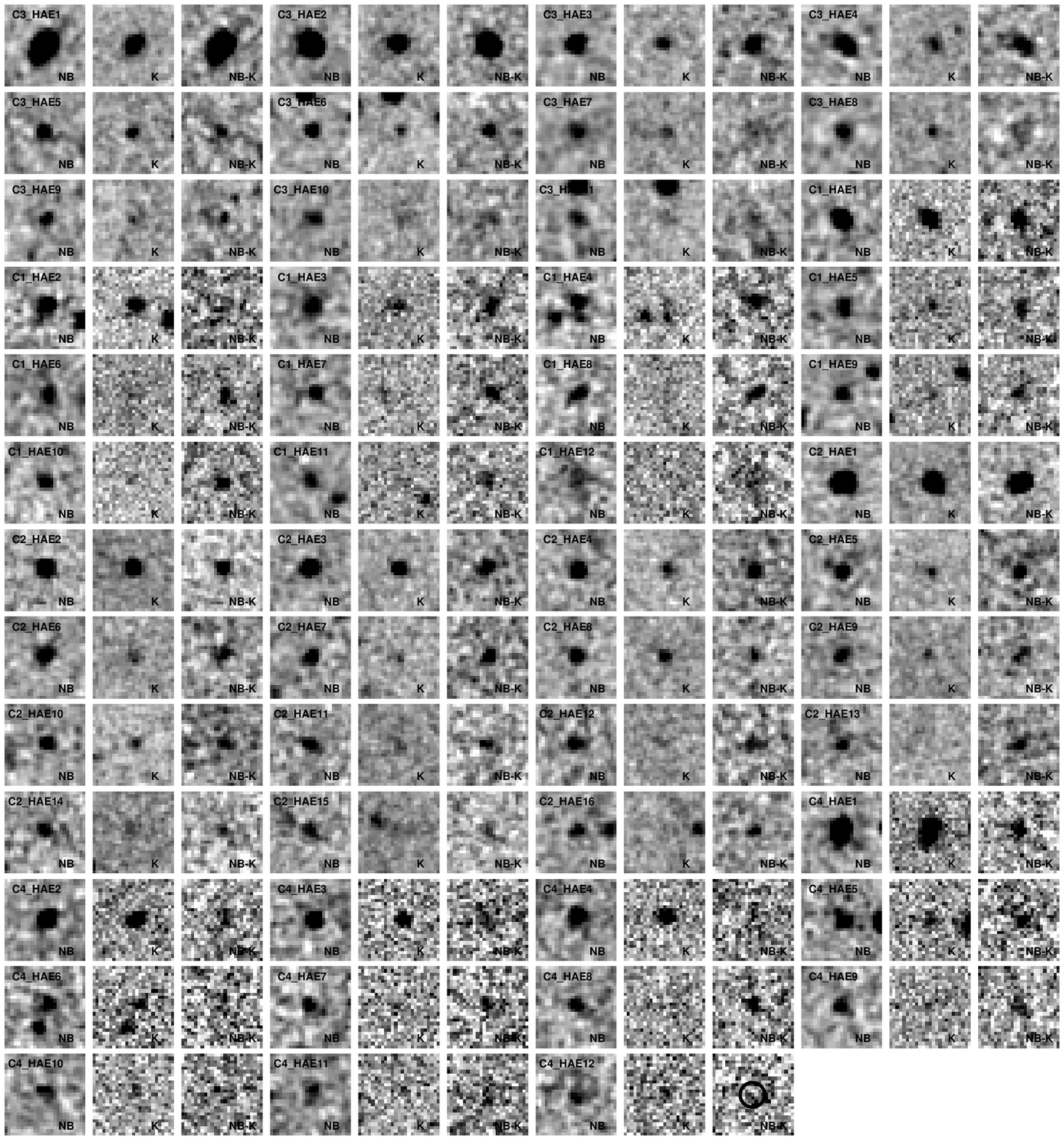}
  \caption{WFCAM thumbnail images of the emitter candidates in the 0200+015 and surrounding control fields. The size of the images is $10 \times 10$ arcsec$^2$ ($\sim 80 \times 80$\,kpc$^2$).  The circle in the bottom-left panel shows the size of the 3\,arcsec aperture for photometry.  The same magnitude range are used to display for all the images.}
\end{figure*}

%
%
\begin{table*}
\centering
\begin{minipage}{168mm}
 \caption{Properties of the emitter candidates in the SSA\,13 (C1) and the control fields (C2, C3, and C4)}
 \begin{tabular}{@{}lcccccccccccccccc}
  \hline
ID & Coordinate (J2000) & ${\rm H}_2{\rm S1}$ & $K$ & $\Sigma$ & $EW_{\rm obs}$ & log $F_{\rm {H}\alpha}^a$ &log $L_{\rm {H}\alpha}^b$ & SFR$_{\rm {H}\alpha}^c$ & M$_*^d$ & Note \\
 & (h:m:s) (d:m:s) & (mag) & (mag) &  & (\AA) & (cgs) & (cgs) &  & \\
\hline
SSA\,13-C1-HAE1 & 13:12:27.80 +42:42:26.6 & 17.76 & 17.99 & 2.6 & 53 & -15.95 & 42.61 & 201 & 11.3 & ... \\
SSA\,13-C1-HAE2 & 13:12:36.58 +42:40:02.5 & 18.26 & 19.04 & 4.5 & 254 & -15.72 & 42.84 & 387 & 10.8 & BzK \\
SSA\,13-C1-HAE3 & 13:12:42.53 +42:41:56.0 & 18.68 & 19.31 & 2.6 & 187 & -15.95 & 42.61 & 201 & 10.7 & BzK \\
SSA\,13-C1-HAE4 & 13:13:07.72 +42:34:41.0 & 18.71 & 19.81 & 3.7 & 445 & -15.80 & 42.76 & 304 & 10.5 & BzK \\
SSA\,13-C1-HAE5 & 13:12:10.36 +42:43:10.6 & 18.81 & 19.64 & 2.8 & 279 & -15.92 & 42.64 & 218 & 10.6 & BzK \\
SSA\,13-C1-HAE6 & 13:11:59.75 +42:42:53.3 & 18.98 & 20.14 & 3.0 & 490 & -15.90 & 42.66 & 231 & 10.4 & BzK \\
\hline
SSA\,13-C2-HAE1 & 13:14:44.24 +42:35:13.7 & 18.84 & 19.94 & 3.3 & 445 & -15.86 & 42.70 & 261 & 10.5 & ... \\
SSA\,13-C2-HAE2 & 13:15:04.66 +42:39:44.0 & 18.92 & 19.83 & 2.7 & 322 & -15.94 & 42.62 & 207 & 10.5 & ... \\
SSA\,13-C2-HAE3 & 13:15:21.66 +42:46:29.0 & 18.97 & 21.11 & 3.9 & 2331 & -15.78 & 42.78 & 325 & 10.0 & ... \\
\hline
SSA\,13-C3-HAE1 & 13:15:03.25 +43:01:42.6 & 18.15 & 19.28 & 6.2 & 462 & -15.58 & 42.98 & 575 & 10.7 & ... \\
SSA\,13-C3-HAE2 & 13:15:15.31 +43:09:52.7 & 18.32 & 18.71 & 2.5 & 100 & -15.97 & 42.59 & 190 & 11.0 & ... \\
SSA\,13-C3-HAE3 & 13:15:13.54 +43:08:30.1 & 18.39 & 18.93 & 3.0 & 150 & -15.89 & 42.67 & 238 & 10.9 & ... \\
SSA\,13-C3-HAE4 & 13:15:21.67 +43:06:01.3 & 18.49 & 19.61 & 4.6 & 457 & -15.71 & 42.85 & 394 & 10.6 & ... \\
SSA\,13-C3-HAE5 & 13:14:22.61 +43:04:44.9 & 18.78 & 19.49 & 2.6 & 219 & -15.96 & 42.61 & 197 & 10.7 & ... \\
\hline
SSA\,13-C4-HAE1 & 13:12:39.90 +43:12:11.9 & 17.24 & 17.63 & 6.7 & 99 & -15.55 & 43.02 & 630 & 11.4 & ... \\
SSA\,13-C4-HAE2 & 13:11:58.12 +43:06:47.3 & 18.24 & 18.71 & 3.1 & 124 & -15.88 & 42.68 & 244 & 11.0 & ... \\
SSA\,13-C4-HAE3 & 13:12:26.80 +43:01:46.4 & 18.86 & 19.65 & 2.6 & 260 & -15.96 & 42.61 & 199 & 10.6 & ... \\
SSA\,13-C4-HAE4 & 13:12:45.44 +43:13:32.7 & 18.89 & 20.39 & 3.7 & 840 & -15.80 & 42.76 & 303 & 10.3 & ... \\
SSA\,13-C4-HAE5 & 13:12:10.09 +43:07:36.8 & 18.89 & 19.94 & 3.0 & 408 & -15.89 & 42.67 & 238 & 10.5 & ... \\
SSA\,13-C4-HAE6 & 13:12:22.13 +43:01:34.0 & 18.98 & 19.99 & 2.7 & 382 & -15.94 & 42.63 & 210 & 10.5 & ... \\
\hline
 \end{tabular}
$^a$ The H$\alpha$ flux (erg s$^{-1}$ cm$^{-2}$) corrected for 33\% [N{\sc ii}] contribution to the measured flux.\\
$^b$ The H$\alpha$ luminosity (erg s$^{-1}$).\\
$^c$ The attenuation corrected star-formation rate derived from the H$\alpha$ luminosity (M$_{\sun}$ yr$^{-1}$).\\
$^d$ The stellar mass estimated from the $K$-band magnitude by assuming a constant mass-to-light ratio from \citet{2008MNRAS.388.1473G}.\\
\end{minipage}
\end{table*}

%
%
\begin{figure*}
\centering
  \includegraphics[scale=.9]{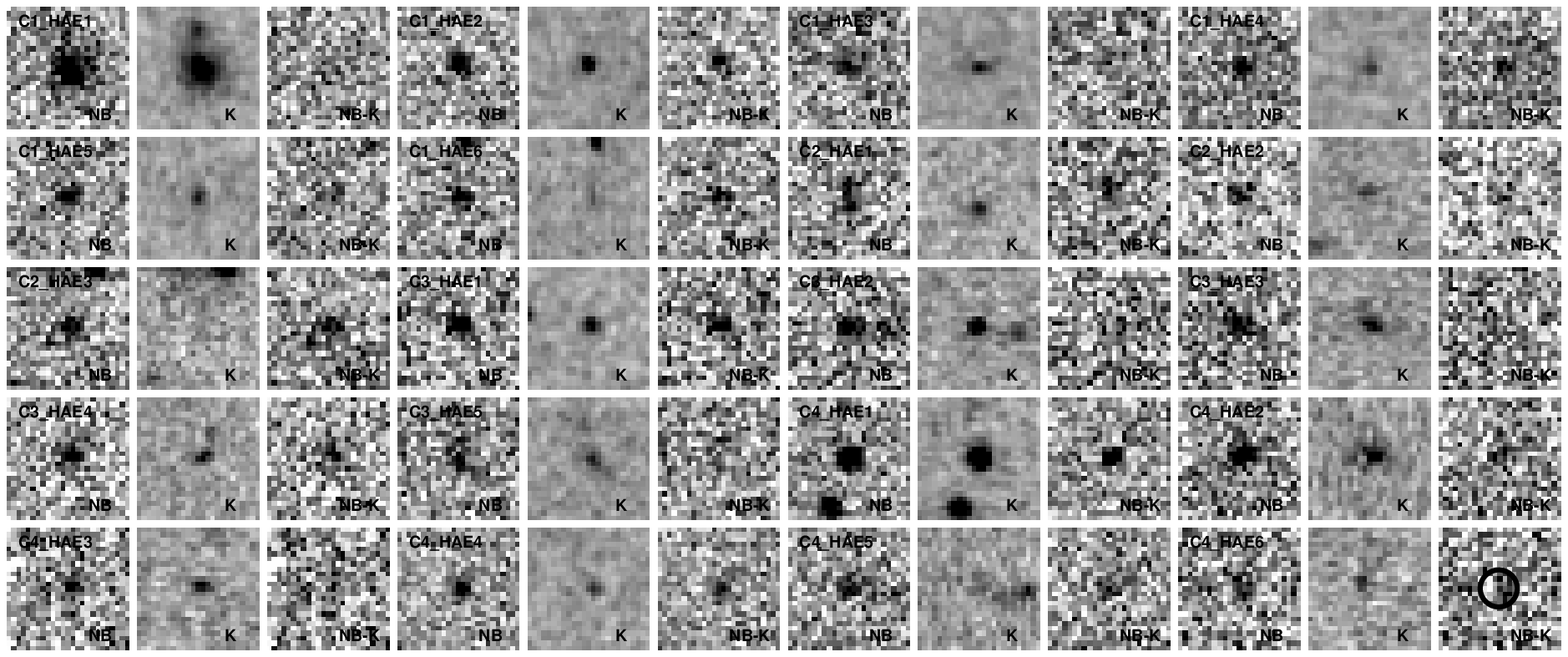}
  \caption{WFCAM thumbnail images of the emitter candidates in the SSA\,13 and surrounding control fields. The size of the images is $10 \times 10$ arcsec$^2$ ($\sim 80 \times 80$\,kpc$^2$).  The circle in the bottom-left panel shows the size of the 3\,arcsec aperture for photometry.  The same magnitude range are used to display for all the images.}
\end{figure*}

\section[]{Full sky maps of emitters}

%
%
\begin{figure*}
\centering
  \includegraphics[scale=0.9]{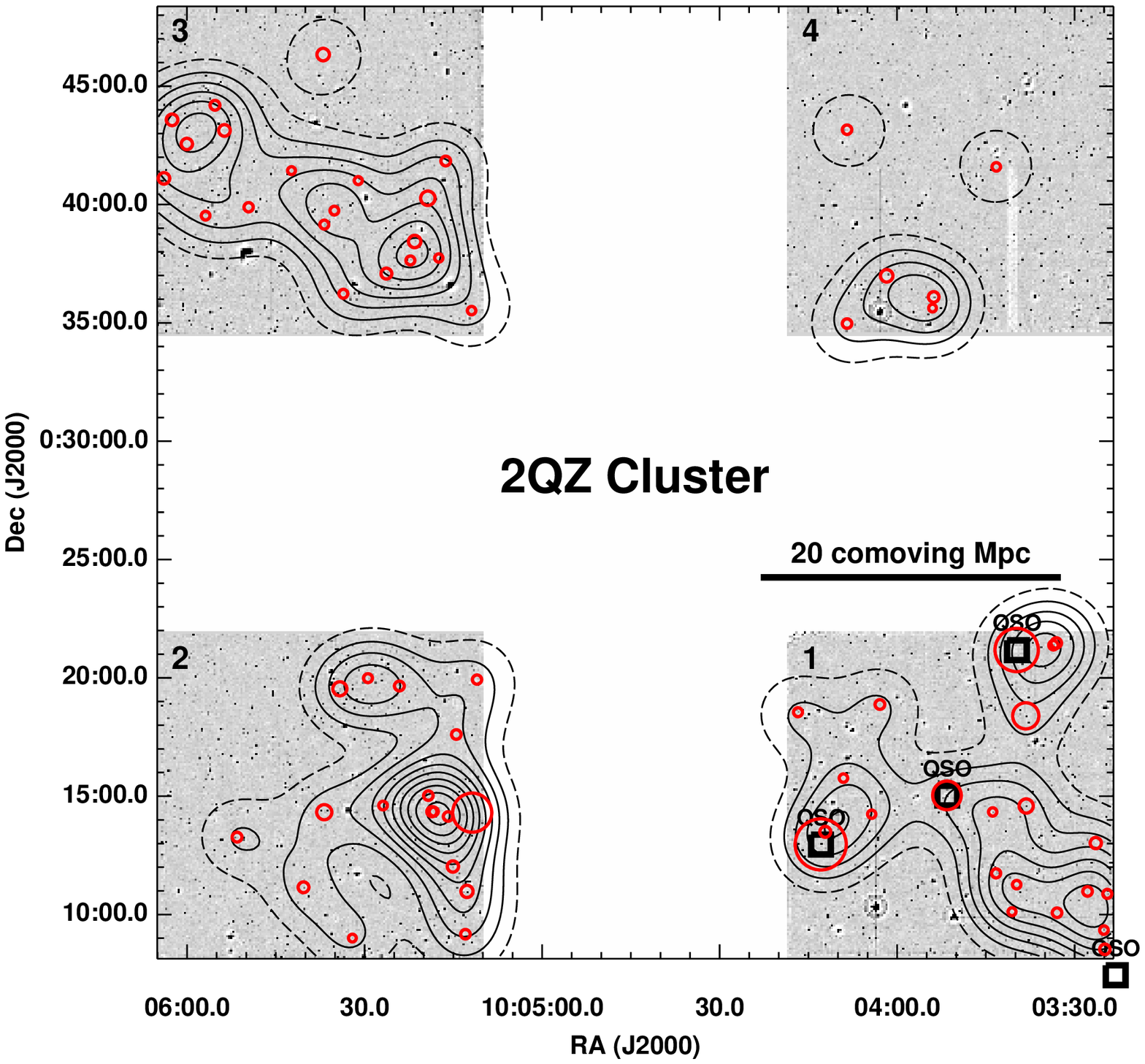}
  \caption{The sky distribution and smoothed density map of emitter candidates in the 2QZ cluster field.  The squares indicate the positions of the four target QSOs.  The circles indicate the emitter candidates and the size is proportional to the emission-line flux.  The thick horizontal bars show the angular scale of 20 comoving\,Mpc (6.2\,Mpc in physical scale) at $z=2.23$. The contours show deviations of local emitter densities from the average densities, from -0.5 to 3\,$\sigma$ with 0.5\,$\sigma$ intervals.}
\end{figure*}

%
%
\begin{figure*}
\centering
  \includegraphics[scale=0.9]{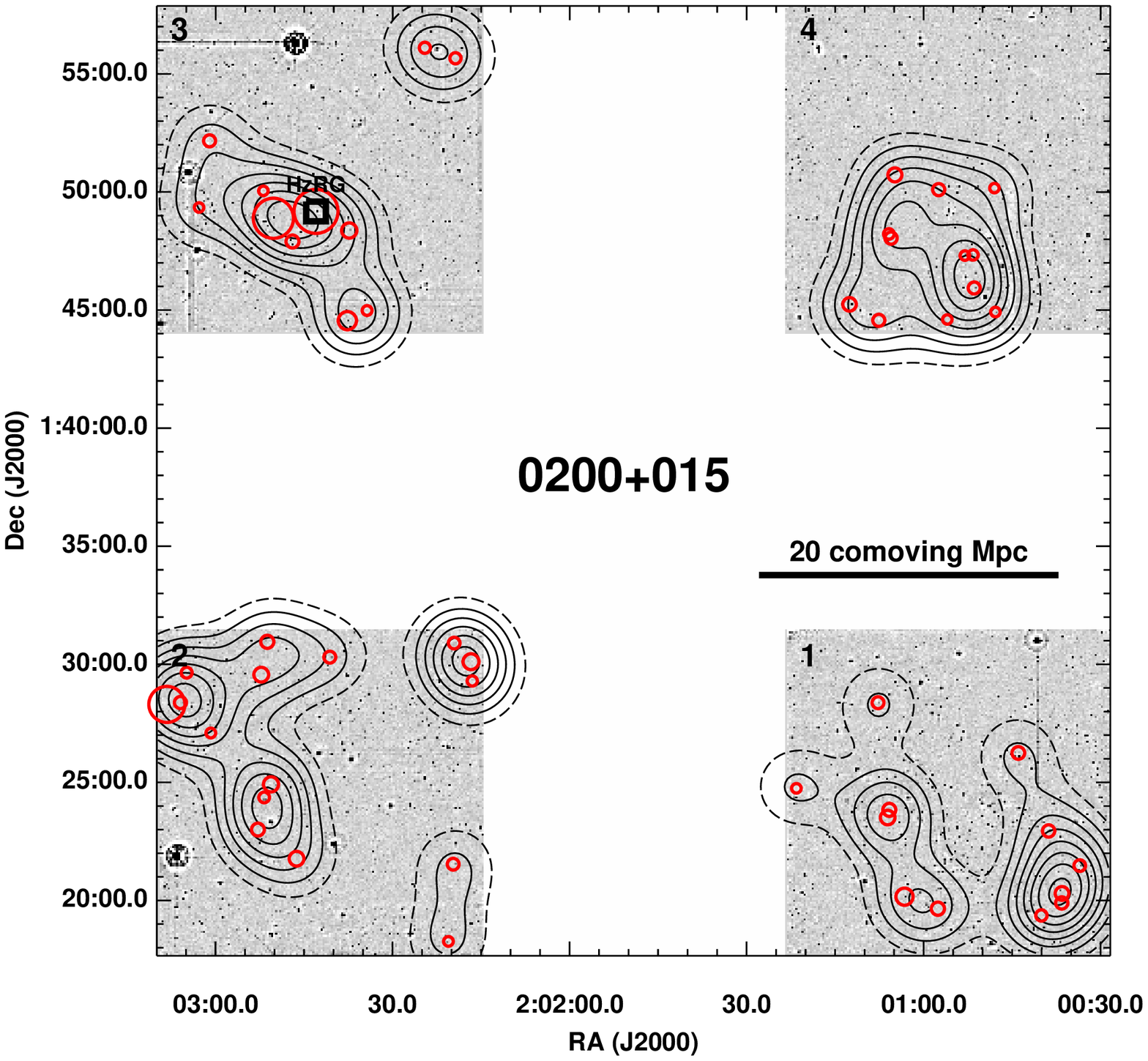}
  \caption{The sky distribution and smoothed density map of emitter candidates the 0200+015 field.  The square indicate the positions of the HzRG.  The circles indicate the emitter candidates and the size is proportional to the emission-line flux.  The thick horizontal bars show the angular scale of 20 comoving\,Mpc (6.2\,Mpc in physical scale) at $z=2.23$. The contours show deviations of local emitter densities from the average densities, from -0.5 to 3\,$\sigma$ with 0.5\,$\sigma$ intervals.}
\end{figure*}

%
%
\begin{figure*}
\centering
  \includegraphics[scale=0.9]{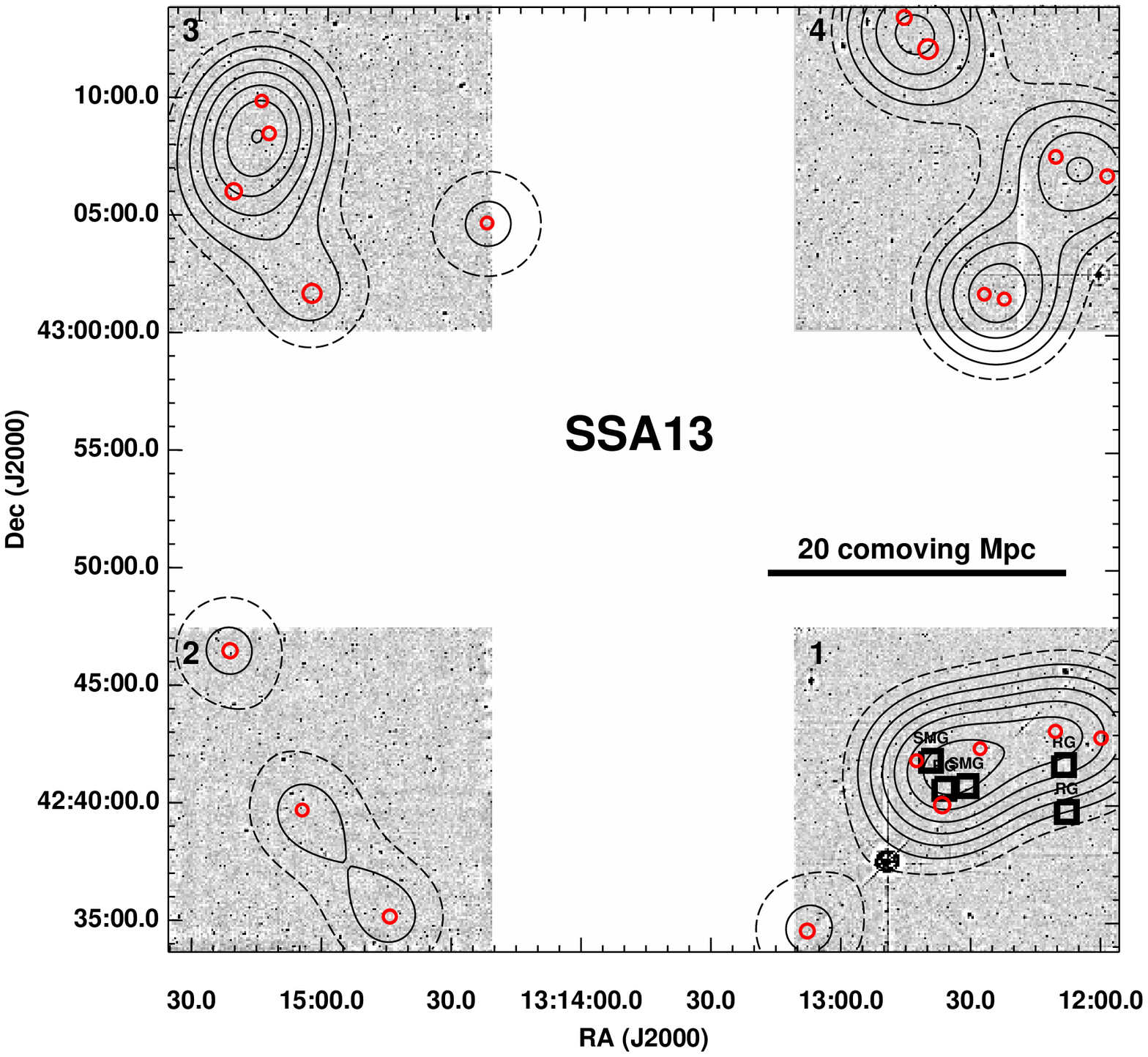}
  \caption{The sky distribution and smoothed density map of emitter candidates at $z\sim 2.23$ in the SSA\,13 field.  The squares indicate the positions of the SMGs and OFRGs. The circles indicate the emitter candidates and the size is proportional to the emission-line flux.  The thick horizontal bars show the angular scale of 20 comoving\,Mpc (6.2\,Mpc in physical scale) at $z=2.23$.  The contours show deviations of local emitter densities from the average densities, from -0.5 to 3\,$\sigma$ with 0.5\,$\sigma$ intervals.}
\end{figure*}

\label{lastpage}


\begin{thebibliography}{99}

\bibitem[\protect\citeauthoryear{Becker, White, \& Helfand}{1995}]{1995ApJ...450..559B} Becker R.~H., White R.~L., Helfand D.~J., 1995, ApJ, 450, 559 

\bibitem[\protect\citeauthoryear{Bertin \& Arnouts}{1996}]{1996A&AS..117..393B} Bertin, E., \& Arnouts, S.\ 1996, A\&A, 117, 393 

\bibitem[\protect\citeauthoryear{Bertin et al.}{2002}]{2002ASPC..281..228B} Bertin E., Mellier Y., Radovich M., Missonnier G., Didelon P., Morin B., 2002, ASPC, 281, 228 

\bibitem[\protect\citeauthoryear{Best et al.}{2003}]{2003MNRAS.343....1B} Best P.~N., Lehnert M.~D., Miley G.~K., R{\"o}ttgering H.~J.~A., 2003, MNRAS, 343, 1 

\bibitem[\protect\citeauthoryear{Blakeslee et al.}{2003}]{2003ApJ...596L.143B} Blakeslee J.~P., et al., 2003, ApJ, 596, L143 

\bibitem[\protect\citeauthoryear{Casali et al.}{2007}]{2007A&A...467..777C} Casali M., et al., 2007, A\&A, 467, 777 

\bibitem[\protect\citeauthoryear{Chapman et al.}{2005}]{2005ApJ...622..772C} Chapman S.~C., Blain A.~W., Smail I., Ivison R.~J., 2005, ApJ, 622, 772 

\bibitem[\protect\citeauthoryear{Chapman et al.}{2009}]{2009ApJ...691..560C} Chapman S.~C., Blain A., Ibata R., Ivison R.~J., Smail I., Morrison G., 2009, ApJ, 691, 560 

\bibitem[\protect\citeauthoryear{Chen, Lanzetta, \& Webb}{2001}]{2001ApJ...556..158C} Chen H.-W., Lanzetta K.~M., Webb J.~K., 2001, ApJ, 556, 158 

\bibitem[\protect\citeauthoryear{Clowes \& Campusano}{1991}]{1991MNRAS.249..218C} Clowes R.~G., Campusano L.~E., 1991, MNRAS, 249, 218 

\bibitem[\protect\citeauthoryear{Condon et al.}{1998}]{1998AJ....115.1693C} Condon J.~J., Cotton W.~D., Greisen E.~W., Yin Q.~F., Perley R.~A., Taylor G.~B., Broderick J.~J., 1998, AJ, 115, 1693 

\bibitem[\protect\citeauthoryear{Crighton et al.}{2010}]{2010arXiv1006.4385C} Crighton N.~H.~M., et al., 2010, arXiv, arXiv:1006.4385 

\bibitem[\protect\citeauthoryear{Croom et al.}{2001}]{2001MNRAS.322L..29C} Croom S.~M., Smith R.~J., Boyle B.~J., Shanks T., Loaring N.~S., Miller L., Lewis I.~J., 2001, MNRAS, 322, L29 

\bibitem[\protect\citeauthoryear{Croom et al.}{2004}]{2004MNRAS.349.1397C} Croom S.~M., Smith R.~J., Boyle B.~J., Shanks T., Miller L., Outram P.~J., Loaring N.~S., 2004, MNRAS, 349, 1397 

\bibitem[\protect\citeauthoryear{Cutri et al.}{2003}]{2003tmc..book.....C} Cutri R.~M., et al., 2003, 2MASS All Sky Catalog of point sources, The IRSA 2MASS All-Sky Point Source Catalog, NASA/IPAC Infrared Science Archive. http://irsa.ipac.caltech.edu/applications/Gator/ 

\bibitem[\protect\citeauthoryear{Daddi et al.}{2004}]{2004ApJ...617..746D} Daddi E., Cimatti A., Renzini A., Fontana A., Mignoli M., Pozzetti L., Tozzi P., Zamorani G., 2004, ApJ, 617, 746 

\bibitem[\protect\citeauthoryear{Digby-North et al.}{2010}]{2010MNRAS.407..846D} Digby-North J.~A., et al., 2010, MNRAS, 407, 846 

\bibitem[\protect\citeauthoryear{Dressler}{1980}]{1980ApJ...236..351D} Dressler A., 1980, ApJ, 236, 351 

\bibitem[\protect\citeauthoryear{Elbaz et al.}{2007}]{2007A&A...468...33E} Elbaz D., et al., 2007, A\&A, 468, 33 

\bibitem[\protect\citeauthoryear{Ellingson, Yee, \& Green}{1991}]{1991ApJ...371...49E} Ellingson E., Yee H.~K.~C., Green R.~F., 1991, ApJ, 371, 49 

\bibitem[\protect\citeauthoryear{Ellis et al.}{1997}]{1997ApJ...483..582E} Ellis R.~S., Smail I., Dressler A., Couch W.~J., Oemler A., Jr., Butcher H., Sharples R.~M., 1997, ApJ, 483, 582 

\bibitem[\protect\citeauthoryear{Garn et al.}{2010}]{2010MNRAS.402.2017G} Garn T., et al., 2010, MNRAS, 402, 2017 

\bibitem[\protect\citeauthoryear{Geach et al.}{2008}]{2008MNRAS.388.1473G} Geach J.~E., Smail I., Best P.~N., Kurk J., Casali M., Ivison R.~J., Coppin K., 2008, MNRAS, 388, 1473 (G08)

\bibitem[\protect\citeauthoryear{Gr{\"u}tzbauch et al.}{2011}]{2011MNRAS.tmp...47G} Gr{\"u}tzbauch R., Chuter R.~W., Conselice C.~J., Bauer A.~E., Bluck A.~F.~L., Buitrago F., Mortlock A., 2011, MNRAS, 47 

\bibitem[\protect\citeauthoryear{G{\'o}mez et al.}{2003}]{2003ApJ...584..210G} G{\'o}mez P.~L., et al., 2003, ApJ, 584, 210 

\bibitem[\protect\citeauthoryear{Hall \& Green}{1998}]{1998ApJ...507..558H} Hall P.~B., Green R.~F., 1998, ApJ, 507, 558 

\bibitem[\protect\citeauthoryear{Hatch et al.}{2011}]{2011arXiv1103.4364H} Hatch N.~A., Kurk J.~D., Pentericci L., Venemans B.~P., Kuiper E., Miley 
G.~K., R{\"o}ttgering H.~J.~A., 2011, arXiv, arXiv:1103.4364 

\bibitem[\protect\citeauthoryear{Hayashi et al.}{2009}]{2009ApJ...691..140H} Hayashi M., et al., 2009, ApJ, 691, 140 

\bibitem[\protect\citeauthoryear{Hayashi et al.}{2010}]{2010MNRAS.402.1980H} Hayashi M., Kodama T., Koyama Y., Tanaka I., Shimasaku K., Okamura S., 2010, MNRAS, 402, 1980 

\bibitem[\protect\citeauthoryear{Hayashino et al.}{2004}]{2004AJ....128.2073H} Hayashino T., et al., 2004, AJ, 128, 2073 

\bibitem[\protect\citeauthoryear{Hayes, Schaerer, {\"O}stlin}{2010}]{2010A&A...509L...5H} Hayes M., Schaerer D., {\"O}stlin G., 2010, A\&A, 509, L5 

\bibitem[\protect\citeauthoryear{Hu \& McMahon}{1996}]{1996Natur.382..231H} Hu E.~M., McMahon R.~G., 1996, Natur, 382, 231 

\bibitem[\protect\citeauthoryear{Iwamuro et al.}{2003}]{2003ApJ...598..178I} Iwamuro F., et al., 2003, ApJ, 598, 178 

\bibitem[\protect\citeauthoryear{Kajisawa et al.}{2006}]{2006MNRAS.371..577K} Kajisawa M., Kodama T., Tanaka I., Yamada T., Bower R., 2006, MNRAS, 371, 577 

\bibitem[\protect\citeauthoryear{Keel et al.}{1999}]{1999AJ....118.2547K} Keel W.~C., Cohen S.~H., Windhorst R.~A., Waddington I., 1999, AJ, 118, 2547 

\bibitem[\protect\citeauthoryear{Kennicutt}{1998}]{1998ARA&A..36..189K} Kennicutt R.~C., Jr., 1998, ARA\&A, 36, 189 

\bibitem[\protect\citeauthoryear{Kashikawa et al.}{2007}]{2007ApJ...663..765K} Kashikawa N., Kitayama T., Doi M., Misawa T., Komiyama Y., Ota K., 2007, ApJ, 663, 765 

\bibitem[\protect\citeauthoryear{Kim et al.}{2011}]{2011MNRAS.410..241K} Kim J.-W., Edge A.~C., Wake D.~A., Stott J.~P., 2011, MNRAS, 410, 241

\bibitem[\protect\citeauthoryear{Kimura et al.}{2010}]{2010PASJ...62.1135K} Kimura M., et al., 2010, PASJ, 62, 1135 

\bibitem[\protect\citeauthoryear{Kodama et al.}{2007}]{2007MNRAS.377.1717K} Kodama T., Tanaka I., Kajisawa M., Kurk J., Venemans B., De Breuck C., Vernet J., Lidman C., 2007, MNRAS, 377, 1717 

\bibitem[\protect\citeauthoryear{Kurk et al.}{2000}]{2000A&A...358L...1K} Kurk J.~D., et al., 2000, A\&A, 358, L1 

\bibitem[\protect\citeauthoryear{Kurk et al.}{2004a}]{2004A&A...428..793K} Kurk J.~D., Pentericci L., R{\"o}ttgering H.~J.~A., Miley G.~K., 2004, A\&A, 428, 793 

\bibitem[\protect\citeauthoryear{Kurk et al.}{2004b}]{2004A&A...428..817K} Kurk J.~D., Pentericci L., Overzier R.~A., R{\"o}ttgering H.~J.~A., Miley G.~K., 2004, A\&A, 428, 817 

\bibitem[\protect\citeauthoryear{Kurk et al.}{2009}]{2009A&A...504..331K} Kurk J., et al., 2009, A\&A, 504, 331 


\bibitem[\protect\citeauthoryear{Labb{\'e} et al.}{2003}]{2003ApJ...591L..95L} Labb{\'e} I., et al., 2003, ApJ, 591, L95 

\bibitem[\protect\citeauthoryear{Landolt}{1992}]{1992AJ....104..340L} Landolt A.~U., 1992, AJ, 104, 340 

\bibitem[\protect\citeauthoryear{Large et al.}{1981}]{1981MNRAS.194..693L} Large M.~I., Mills B.~Y., Little A.~G., Crawford D.~F., Sutton J.~M., 1981, MNRAS, 194, 693 

\bibitem[\protect\citeauthoryear{Lehmer et al.}{2009}]{2009ApJ...691..687L} Lehmer B.~D., et al., 2009, ApJ, 691, 687 

\bibitem[\protect\citeauthoryear{Lewis et al.}{2002}]{2002MNRAS.334..673L} Lewis I., et al., 2002, MNRAS, 334, 673 

\bibitem[\protect\citeauthoryear{Lilly et al.}{1995}]{1995ApJ...455..108L} Lilly S.~J., Tresse L., Hammer F., Crampton D., Le Fevre O., 1995, ApJ, 455, 108 

\bibitem[\protect\citeauthoryear{Matsuda et al.}{2004}]{2004AJ....128..569M} Matsuda Y., et al., 2004, AJ, 128, 569 

\bibitem[\protect\citeauthoryear{Matsuda et al.}{2005}]{2005ApJ...634L.125M} Matsuda Y., et al., 2005, ApJ, 634, L125 

\bibitem[\protect\citeauthoryear{Matsuda et al.}{2009}]{2009MNRAS.400L..66M} Matsuda Y., et al., 2009, MNRAS, 400, L66 

\bibitem[\protect\citeauthoryear{Matsuda et al.}{2010}]{2010MNRAS.403L..54M} Matsuda Y., et al., 2010, MNRAS, 403, L54 

\bibitem[\protect\citeauthoryear{Matsuda et al.}{2011}]{2011MNRAS.410L..13M} Matsuda Y., et al., 2011, MNRAS, 410, L13 

\bibitem[\protect\citeauthoryear{McCarthy, Elston, \& Eisenhardt}{1992}]{1992ApJ...387L..29M} McCarthy P.~J., Elston R., Eisenhardt P., 1992, ApJ, 387, L29 

\bibitem[\protect\citeauthoryear{McCarthy}{1993}]{1993ARA&A..31..639M} McCarthy P.~J., 1993, ARA\&A, 31, 639 

\bibitem[\protect\citeauthoryear{Mei et al.}{2009}]{2009ApJ...690...42M} Mei S., et al., 2009, ApJ, 690, 42 

\bibitem[\protect\citeauthoryear{Miley et al.}{2004}]{2004Natur.427...47M} Miley G.~K., et al., 2004, Natur, 427, 47 

\bibitem[\protect\citeauthoryear{Miley \& De Breuck}{2007}]{2007A&ARv..15...67M} Miley G., De Breuck C., 2007, A\&ARv, 15, 67 

\bibitem[\protect\citeauthoryear{Miyazaki et al.}{2002}]{2002PASJ...54..833M} Miyazaki S., et al., 2002, PASJ, 54, 833

\bibitem[\protect\citeauthoryear{Oke et al.}{1995}]{1995PASP..107..375O} Oke J.~B., et al., 1995, PASP, 107, 375 

\bibitem[\protect\citeauthoryear{Ouchi et al.}{2004}]{2004ApJ...611..660O} Ouchi M., et al., 2004, ApJ, 611, 660 


\bibitem[\protect\citeauthoryear{Palunas et al.}{2004}]{2004ApJ...602..545P} Palunas P., Teplitz H.~I., Francis P.~J., Williger G.~M., Woodgate B.~E., 2004, ApJ, 602, 545 

\bibitem[\protect\citeauthoryear{Pascarelle et al.}{1996}]{1996Natur.383...45P} Pascarelle S.~M., Windhorst R.~A., Keel W.~C., Odewahn S.~C., 1996, Natur, 383, 45 

\bibitem[\protect\citeauthoryear{Pentericci et al.}{2000}]{2000A&A...361L..25P} Pentericci L., et al., 2000, A\&A, 361, L25 

\bibitem[\protect\citeauthoryear{R{\"o}ttgering et al.}{1997}]{1997A&A...326..505R} R{\"o}ttgering H.~J.~A., van Ojik R., Miley G.~K., Chambers K.~C., van Breugel W.~J.~M., de Koff S., 1997, A\&A, 326, 505 

\bibitem[\protect\citeauthoryear{Schlegel, Finkbeiner, \& Davis}{1998}]{1998ApJ...500..525S} Schlegel D.~J., Finkbeiner D.~P., Davis M., 1998, ApJ, 500, 525 

\bibitem[\protect\citeauthoryear{Shen et al.}{2007}]{2007AJ....133.2222S} Shen Y., et al., 2007, AJ, 133, 2222 

\bibitem[\protect\citeauthoryear{Smail et al.}{2003}]{2003ApJ...599...86S} Smail I., Scharf C.~A., Ivison R.~J., Stevens J.~A., Bower R.~G., Dunlop J.~S., 2003, ApJ, 599, 86 

\bibitem[\protect\citeauthoryear{Smail et al.}{2004}]{2004ApJ...616...71S} Smail I., Chapman S.~C., Blain A.~W., Ivison R.~J., 2004, ApJ, 616, 71 

\bibitem[\protect\citeauthoryear{Smith et al.}{2002}]{2002AJ....123.2121S} Smith J.~A., et al., 2002, AJ, 123, 2121 

\bibitem[\protect\citeauthoryear{Sobral et al.}{2011}]{2011MNRAS.411..675S} Sobral D., Best P.~N., Smail I., Geach J.~E., Cirasuolo M., Garn T., Dalton G.~B., 2011, MNRAS, 411, 675 

\bibitem[\protect\citeauthoryear{Sobral et al.}{2010}]{2010MNRAS.404.1551S} Sobral D., Best P.~N., Geach J.~E., Smail I., Cirasuolo M., Garn T., Dalton G.~B., Kurk J., 2010, MNRAS, 404, 1551 

\bibitem[\protect\citeauthoryear{Sobral et al.}{2009b}]{2009MNRAS.398L..68S} Sobral D., et al., 2009, MNRAS, 398, L68 

\bibitem[\protect\citeauthoryear{Sobral et al.}{2009a}]{2009MNRAS.398...75S} Sobral D., et al., 2009, MNRAS, 398, 75 

\bibitem[\protect\citeauthoryear{Swinbank et al.}{2004}]{2004ApJ...617...64S} Swinbank A.~M., Smail I., Chapman S.~C., Blain A.~W., Ivison R.~J., Keel W.~C., 2004, ApJ, 617, 64 

\bibitem[\protect\citeauthoryear{Steidel et al.}{1998}]{1998ApJ...492..428S} Steidel C.~C., Adelberger K.~L., Dickinson M., Giavalisco M., Pettini M., Kellogg M., 1998, ApJ, 492, 428 

\bibitem[\protect\citeauthoryear{Steidel et al.}{2000}]{2000ApJ...532..170S} Steidel, C.~C., Adelberger, K.~L., Shapley, A.~E., Pettini, M., Dickinson, M., \& Giavalisco, M.\ 2000, ApJ, 532, 170 

\bibitem[\protect\citeauthoryear{Steidel et al.}{2005}]{2005ApJ...626...44S} Steidel C.~C., Adelberger K.~L., Shapley A.~E., Erb D.~K., Reddy N.~A., Pettini M., 2005, ApJ, 626, 44 

\bibitem[\protect\citeauthoryear{Steidel et al.}{2011}]{2011arXiv1101.2204S} Steidel C.~C., Bogosavljevi{\'c} M., Shapley A.~E., Kollmeier J.~A., Reddy N.~A., Erb D.~K., Pettini M., 2011, arXiv, arXiv:1101.2204 

\bibitem[\protect\citeauthoryear{Stevens et al.}{2003}]{2003Natur.425..264S} Stevens J.~A., et al., 2003, Natur, 425, 264 

\bibitem[\protect\citeauthoryear{Suzuki et al.}{2008}]{2008PASJ...60.1347S} Suzuki R., et al., 2008, PASJ, 60, 1347 

\bibitem[\protect\citeauthoryear{Tadaki et al.}{2010}]{2010arXiv1012.4860T} Tadaki K.-i., Kodama T., Koyama Y., Hayashi M., Tanaka I., Tokoku C., 2010, arXiv, arXiv:1012.4860 

\bibitem[\protect\citeauthoryear{Tanaka et al.}{2000}]{2000ApJ...528..123T} Tanaka I., Yamada T., Arag{\'o}n-Salamanca A., Kodama T., Miyaji T., Ohta K., Arimoto N., 2000, ApJ, 528, 123 

\bibitem[\protect\citeauthoryear{Tanaka et al.}{2001}]{2001ApJ...547..521T} Tanaka I., Yamada T., Turner E.~L., Suto Y., 2001, ApJ, 547, 521 

\bibitem[\protect\citeauthoryear{Tanaka et al.}{2010}]{2010arXiv1012.1869T} Tanaka I., et al., 2010, arXiv, arXiv:1012.1869 

\bibitem[\protect\citeauthoryear{Tanaka et al.}{2010}]{2010A&A...518A..18T} Tanaka M., De Breuck C., Venemans B., Kurk J., 2010, A\&A, 518, A18 

\bibitem[\protect\citeauthoryear{Thomas et al.}{2005}]{2005ApJ...621..673T} Thomas D., Maraston C., Bender R., Mendes de Oliveira C., 2005, ApJ, 621, 673 

\bibitem[\protect\citeauthoryear{Tran et al.}{2010}]{2010ApJ...719L.126T} Tran K.-V.~H., et al., 2010, ApJ, 719, L126 

\bibitem[\protect\citeauthoryear{van der Werf, Moorwood, \& Bremer}{2000}]{2000A&A...362..509V} van der Werf P.~P., Moorwood A.~F.~M., Bremer M.~N., 2000, A\&A, 362, 509 

\bibitem[\protect\citeauthoryear{Venemans et al.}{2007}]{2007A&A...461..823V} Venemans B.~P., et al., 2007, A\&A, 461, 823 

\bibitem[\protect\citeauthoryear{Wilman et al.}{2007}]{2007MNRAS.375..735W} Wilman R.~J., Morris S.~L., Jannuzi B.~T., Dav{\'e} R., Shone A.~M., 2007, MNRAS, 375, 735 

\bibitem[\protect\citeauthoryear{Yagi et al.}{2002}]{2002AJ....123...66Y} Yagi M., Kashikawa N., Sekiguchi M., Doi M., Yasuda N., Shimasaku K., Okamura S., 2002, AJ, 123, 66 

\bibitem[\protect\citeauthoryear{Yang et al.}{2010}]{2010ApJ...719.1654Y} Yang Y., Zabludoff A., Eisenstein D., Dav{\'e} R., 2010, ApJ, 719, 1654 

\bibitem[\protect\citeauthoryear{Young, Sargent, \& Boksenberg}{1982}]{1982ApJS...48..455Y} Young P., Sargent W.~L.~W., Boksenberg A., 1982, ApJS, 48, 455 

\end{thebibliography}
\end{document}